\documentstyle[12pt]{article}

\advance \topmargin by -\headheight
\advance \topmargin by -\headsep

\evensidemargin \oddsidemargin
\def\ie{{\em i.e.}}
\def\ie{\hbox{\it i.e.}}

\def\CC{{\mathchoice
{\rm C\mkern-8mu\vrule height1.45ex depth-.05ex
width.05em\mkern9mu\kern-.05em}
{\rm C\mkern-8mu\vrule height1.45ex depth-.05ex
width.05em\mkern9mu\kern-.05em}
{\rm C\mkern-8mu\vrule height1ex depth-.07ex
width.035em\mkern9mu\kern-.035em}
{\rm C\mkern-8mu\vrule height.65ex depth-.1ex
width.025em\mkern8mu\kern-.025em}}}

\def\RR{{\rm I\kern-1.6pt {\rm R}}}

\def\ZZ{{\rm Z}\kern-3.8pt {\rm Z} \kern2pt}
\def\IB{\relax{\rm I\kern-.18em B}}
\def\ID{\relax{\rm I\kern-.18em D}}
\def\II{\relax{\rm I\kern-.18em I}}
\def\IP{\relax{\rm I\kern-.18em P}}

\def\np{Nucl. Phys.}
\def\pl{Phys. Lett.}
\def\prl{Phys. Rev. Lett.}
\def\pr{Phys. Rev.}

\def\jhep{J. High Energy Phys.}

\def\cqg{Class. Quant. Grav.}

\newcommand{\beq}{\begin{equation}}
\newcommand{\eeq}{\end{equation}}
\newcommand{\rc}{\nonumber\\}
\newcommand{\bear}{\begin{eqnarray}}
\newcommand{\eear}{\end{eqnarray}}

\def\beqa{\begin{eqnarray}}
\def\eeqa{\end{eqnarray}}

\newfont{\namefont}{cmr10}
\newfont{\addfont}{cmti7 scaled 1440}
\newfont{\boldmathfont}{cmbx10}
\newfont{\headfontb}{cmbx10 scaled 1728}
\renewcommand{\theequation}{{\rm\thesection.\arabic{equation}}}
\begin{document}
\begin{titlepage}

\begin{center} \Large \bf Supersymmetric Intersections of M-branes and pp-waves

\end{center}

\vskip 0.3truein
\begin{center}
J. Mas
\footnote{javmas@usc.es}
and
A.V. Ramallo
\footnote{alfonso@fpaxp1.usc.es}

\vspace{0.3in}

Departamento de F\'\i sica de Part\'\i culas, Universidad de
Santiago de Compostela \\
E-15782 Santiago de Compostela, Spain
\vspace{0.3in}

\end{center}
\vskip 2truein

\begin{center}
\bf ABSTRACT
\end{center}

We study supersymmetric intersections of M2 and M5 branes with different
pp-waves of M-theory. We consider first M-brane probes in the background of
pp-waves and determine under which conditions the embedding is supersymmetric.
We particularize our formalism to the case of pp-waves with 32, 24 and 20
supersymmetries. We  also construct supergravity solutions for the brane-wave
system. Generically these solutions are delocalised along some directions
transverse to the brane and preserve the same number of supersymmetries as in
the brane probe approach.

\vskip2.6truecm
\leftline{US-FT-1/03\hfill March 2003}
\leftline{hep-th/0303193}
\smallskip
\end{titlepage}
\setcounter{footnote}{0}


\setcounter{equation}{0}
\section{Introduction}
\medskip

With the advent of the AdS/CTF duality, the understanding of string
theory with Ramond-Ramond backgrounds has become a subject of great interest
\cite{jm}. Indeed, in order to extend the gauge theory/gravity
correspondence to
the regime in which the gauge coupling is small, one must quantize string
theory
in such backgrounds. Remarkably, there exists a background of the type IIB
theory with a Ramond-Ramond flux, the maximally supersymmetric
pp-wave \cite{Figueroa}, in which superstring theory is exactly solvable
\cite{Metsaev}. This supergravity solution can be obtained \cite{Figueroados}
by performing the Penrose limit  of the $AdS_5\times S^5$
geometry \cite{Penrose, Guven}. This fact opens the possibility of studying the
string theory/ gauge theory correspondence at the level of full string theory
\cite{BMN}.

The study of D-branes in the pp-wave background is obviously interesting in
order to have an insight on the non-perturbative phenomena of string theory in
this geometry and, through the AdS/CFT correspondence, of its gauge theory
dual.
There are at least three ways to study D-branes in the pp-wave background. The
first one is just the original Polchinski approach adapted to this case, \ie\
one studies open strings with Dirichlet boundary conditions which preserve some
amount of supersymmetry. This is the point of view adopted in ref. \cite{DP}.
The second approach to this problem is the brane probe formalism
\cite{Sken}, in
which one considers the Dirac-Born-Infeld action for the D-brane and
looks for solutions of the equation of motion which are invariant under kappa
symmetry. Finally, one can, as in refs. \cite{KNS, CLP, Bain} , try to find
supergravity solutions representing the intersection between the D-brane
and the
pp-wave. In this case the number of supersymmetries preserved by the
configuration is just the number of Killing spinors of the supergravity
solution
representing the pp-wave/D-brane intersection.

In eleven dimensional supergravity the maximally supersymmetric pp-wave
solution was found long time ago in ref. \cite{KG}. As in the ten dimensional
case, this pp-wave with 32 supersymmetries can be obtained by means of a
Penrose limit of the $AdS_4\times S^7$ and $AdS_7\times S^4$ solutions
\cite{Figueroados}. Actually, there exist some pp-wave backgrounds which, in
addition to the 16 standard supersymmetries preserved by a generic pp-wave,
 are also invariant under a set of supersymmetry transformations along some
so-called ``supernumerary" Killing spinors \cite{CLPdos, GH}. These backgrounds
can be obtained in some cases as Penrose limits of (non-standard) brane
intersections \cite{CLPdos,nonstandard, Singh}. The matrix theory for these
M-theory pp-waves was proposed in ref.
\cite{BMN} and is usually referred to as the BMN matrix theory. As compared
with
the original matrix theory, the BMN matrix action contains mass terms, both for
bosons and fermions, as well as a cubic interaction term, the so-called Myers
term. This action can be obtained from a matrix regularization of the
supermembrane action in the pp-wave geometry \cite{DSJR,SY}.

On can use the BMN matrix theory to find BPS objects on the pp-wave. This is
the approach followed in refs.
\cite{SY}-\cite{Lee}.
Here we will adopt a different point of view
to deal with the problem of finding supersymmetric intersections with the
M-theory pp-waves. We will consider first M2 and M5 brane probes in the pp-wave
background. Although our formalism is valid for a general case, we will mostly
concentrate our analysis in the pp-wave backgrounds which preserve 32, 24 and
20 supersymmetries. The basic tool in the brane probe approach is kappa
symmetry, which provides a condition to be satisfied by the Killing spinors if
the corresponding brane embedding is to be supersymmetric.

For a given M2 and
M5 brane embedding, the number of Killing spinors of the background satisfying
the kappa symmetry condition is just the number of supersymmetries of the
brane-wave intersection. Actually, we will restrict ourselves to branes
extended along the two light-cone directions and along some fixed transverse
hyperplane. The corresponding kappa symmetry matrix is just the antisymmetrized
product of constant gamma matrices and the requirements of kappa symmetry
reduce to a set of algebraic constraints to be satisfied by the Killing
spinors. It is not difficult to perform a case by case analysis of these
constraints and determine the number of supersymmetries of each possible
configuration. Actually, as we will see in the explicit examples, the number of
supersymmetries depends on whether the brane is located at the origin in the
transverse space or at an arbitrary point. In general, some supersymmetries are
lost when we move away from the origin in a generic direction. It is also
interesting to determine how many supernumerary Killing spinors of the pp-wave
survive in the intersection with the branes. Generically, the wave-brane
intersections are not invariant under supersymmetries along supernumerary
Killing spinors (specially for branes located at arbitrary points in transverse
space), although there are some distinguished cases in which this does not
occur.

We will also try to find supergravity solutions  representing the brane-wave
intersection. The natural ansatz for the metric of these solutions is
obtained
by including the corresponding warp factors along the directions parallel and
transverse to the brane. These warp factors are powers of a single harmonic
function, which depends on the coordinates transverse to the brane. In
addition,
we expect to have some modifications to the original quadratic profile of the
pp-wave, due to the back-reaction of the brane \cite{Bain}. On the other hand,
the four-form field strength for these solutions is the sum of the constant
flux
corresponding to the brane and the standard M2 or M5 ans\"atze, the latter
being
given in terms of the derivatives of the harmonic function.  Actually, in most
of the cases, the equations of motion of the four-form are satisfied if the M2
or M5 are delocalised along some of their external coordinates, which implies
that the harmonic function only depends on a subset of the external
coordinates.
Once the harmonic function is determined, the profile for the metric of the
intersection can be found by integrating a second-order differential equation,
which is obtained from the Einstein equations.

The analysis of the supersymmetry preserved by our supergravity solutions leads
to a series of conditions for the Killing spinors, which include, in
particular, the algebraic equations found in the brane probe approach. Thus,
only those embeddings which preserve some supersymmetry in the brane probe
approach  can give rise to supersymmetric solutions of the supergravity
equations. In addition, we will get some extra conditions which are a
consequence of the warping of the metric and involve derivatives with
respect to
the external coordinates. These extra conditions fail to be satisfied by some
of the brane embeddings which were found to preserve some supersymmetries in
the brane probe approach.  In the generic situation, however, we find agreement
between the number of supersymmetries obtained in the supergravity
analysis and
the brane probe approach for a brane located at an arbitrary transverse
position.

This paper is organized as follows. In section 2 we first present the general
conditions that make a brane probe supersymmetric in the pp-wave background.
We then apply this general formalism to the case of the maximally
supersymmetric
pp-wave, which was previosly considered in \cite{Kim}, and to the pp-wave
with 24
supersymmetries. In section 3 we introduce our ansatz for the supergravity
solutions corresponding to the wave-brane intersections and discuss their
supersymmetries. The values of the different components of the Ricci tensor for
these metrics are given in appendix A, while the solution of a differential
equation, which appears in the determination of the profile, is worked out in
appendix B. In section 4 we apply our general formalism to the study of
supergravity solutions corresponding to M2 and M5 branes intersecting a
maximally supersymmetric pp-wave, whereas in section 5 the  pp-wave with 24
supersymmetries is considered. The case of the pp-wave with 20 supersymmetries
is treated in appendix C. Finally, in section 6 we summarize our results and
draw some conclusions.

\setcounter{equation}{0}
\section{Probe analysis}
\medskip
Let us consider an eleven dimension pp-wave metric of the type:
\beq
ds_{11}^2\,=\,2dx^{+}\,dx^{-}\,+\,W\,(\,dx^{+}\,)^2\,+\,(\,dx^{i}\,)^2\,\,,
\label{uno}
\eeq
where $(x^+, x^-)$ are light-cone coordinates and the $x^i$'s will be referred
to as transverse coordinates. The function $W$ is the so-called profile of the
pp-wave and we will assume that only depends on the transverse coordinates
$x^i$. On the other hand, the four-form field strength of eleven dimensional
supergravity will be taken as:
\beq
F\,=\,dx^+\wedge\Theta\,\,,
\label{dos}
\eeq
with $\Theta$ being:
\beq
\Theta\,=\,{1\over 6}\,\theta_{ijk}\,dx^i\wedge dx^j\wedge dx^k\,\,.
\label{tres}
\eeq
This  configuration is a solution of Einstein equations if the profile
$W$ satisfies:
\beq
\partial^2_i\,W\,=\,-{1\over 6}\,\theta_{mnl}\,\theta^{mnl}\,\,.
\label{cuatro}
\eeq
Any solution of the above equation gives rise to a background with 16
supersymmetries. However, for some choices of $\Theta$ and $W$ one can have
solutions with more supersymmetry. In this paper we will restrict ourselves to
the case in which $\Theta$ is  given by a four parameter ansatz
\cite{CLPdos,GH} of the type\footnote{There also exists a seven parameter
ansatz \cite{CLPdos,GH} which, in particular, gives rise to an eleven
dimensional pp-wave with 26  supersymmetries \cite{Michelson}.}:
\beq
\Theta=\mu_1dx^1\wedge dx^2\wedge dx^9+
\mu_2dx^3\wedge dx^4\wedge dx^9+
\mu_3dx^5\wedge dx^6\wedge dx^9+
\mu_4dx^7\wedge dx^8\wedge dx^9,
\label{cinco}
\eeq
with the $\mu_i$'s being constants. It follows from eq. (\ref{cuatro})
that $W$ must be  a quadratic function of the transverse coordinates $x^i$.
Actually, if we write:
\beq
W\,=\,-\sum_i \lambda_i^2\,\, (\,x^i\,)^2\,\,.
\label{seis}
\eeq
Then, a solution of the Einstein equations which preserves at least 18
supersymmetries is obtained when the $\lambda$'s and the $\mu$'s are
related as \cite{CLPdos,GH}:
\bear
&&\lambda_1^2\,=\,\lambda_2^2\,=\,{1\over 36}\,
\Big(\,2\mu_1-\mu_2-\mu_3-\mu_4\,\Big)^2\,\,,\rc\rc
&&\lambda_3^2\,=\,\lambda_4^2\,=\,{1\over 36}\,
\Big(\,-\mu_1+2\mu_2-\mu_3-\mu_4\,\Big)^2\,\,,\rc\rc
&&\lambda_5^2\,=\,\lambda_6^2\,=\,{1\over 36}\,
\Big(\,-\mu_1-\mu_2+2\mu_3-\mu_4\,\Big)^2\,\,,\rc\rc
&&\lambda_7^2\,=\,\lambda_8^2\,=\,{1\over 36}\,
\Big(\,-\mu_1-\mu_2-\mu_3+2\mu_4\,\Big)^2\,\,,\rc\rc
&&\lambda_9^2\,=\,{1\over 9}\,
\Big(\,\mu_1+\mu_2+\mu_3+\mu_4\,\Big)^2\,\,,
\label{siete}
\eear

In the study of the supersymmetry of  brane probes in the above geometry
we must
know the  Killing spinors of the background. In order to write their general
form, let us define the matrix:
\beq
\theta\,=\,{1\over 6}\,\,
\theta_{\,\widehat i \,\,\,\widehat j\, \,\,\widehat k}\,\,\,
\Gamma^{\,\widehat i \,\,\,\widehat j\, \,\,\widehat k}\,\,.
\label{ocho}
\eeq
In eq. (\ref{ocho}), and in what follows, hatted indices denote flat components
with respect to the basis of one-forms given in appendix A (see eq.
(\ref{apados})). Then
\cite{Figueroatres}, the Killing spinors $\epsilon$ take the form:
\beq
\epsilon\,=\,(\,1\,+\,x^i\,\Omega_i)\,e^{x^{+}\,\Omega_{+}}\,\chi\,\,,
\label{nueve}
\eeq
where $\chi$ is a constant spinor and the dependence of $\epsilon$ on the
coordinates $x^+$ and $x^i$ is determined by the action of the matrices
$\Omega_{+}$ and $\Omega_i$ on $\chi$. These matrices are given in terms of
$\theta$ as follows:
\beq
\Omega_{+}\,=\,-{1\over 12}\,\theta\,\Big[\,
\Gamma_{\,\widehat -}\,\Gamma_{\,\widehat +}\,+\,1\,\Big]\,,
\,\,\,\,\,\,\,\,\,\,\,\,\,\,\,\,\,\,
\Omega_{i}\,=\,{1\over 24}\,\Big[\,3\theta\Gamma_{\,\widehat i}\,+\,
\Gamma_{\,\widehat i}\,\theta\,\Big]\,\Gamma_{\,\widehat -}\,\,.
\label{diez}
\eeq
The spinor $\chi$ is, in general, not arbitrary but determined by some
algebraic constraints. In particular, there are always 16 Killing spinors,
obtained by solving the equation $\Gamma_{\,\widehat -}\,\chi=0$, which are
the so-called standard spinors. Notice that these standard Killing spinors do
not depend on the transverse coordinates $x^i$ and they can only depend on the
light-cone coordinate $x^+$. The spinors $\epsilon$ for which
$\Gamma_{\,\widehat -}\,\chi\not= 0$ are called supernumerary Killing spinors.
For the background we have written above there are at least two of them. These
supernumerary spinors have a nontrivial dependence on the transverse
coordinates
$x^i$.

Let us now place an M-brane probe in the above pp-wave background. The
supersymmetry preserved by the probe is determined  by the solutions of the
equation:
\beq
\Gamma_{\kappa}\,\epsilon\,=\,\epsilon\,\,,
\label{once}
\eeq
where $\Gamma_{\kappa}$ is the so-called kappa symmetry matrix of the brane
probe,  $\epsilon$ is a Killing spinor of the background and it should be
understood that both sides of this equation are evaluated on
the worldvolume of the brane.

The kappa symmetry matrix for an M2-brane is:

\beq
\Gamma_{\kappa}^{M2}\,=\,{1\over 3!\sqrt{-det g}}\,\,\epsilon^{\mu\nu\rho}\,
\gamma_{\mu\nu\rho}\,\,,
\label{doce}
\eeq
where $g$ is the determinant of the induced metric,
$\gamma_{\mu}\,=\,\partial_{\mu} x^{M}\, E^{\,\widehat  P}_{M}\,
\Gamma_{\,\widehat P}$ are the induced $\Gamma$-matrices, with
$E^{\,\widehat  P}_{M}$ being the vierbeins of the eleven dimensional metric
$G_{MN}$, defined as:
\beq
G_{MN}\,=\,E_{M}^{\,\widehat P}\,E_{N}^{\,\widehat Q}\,
\eta_{\,\widehat P \,\widehat Q}\,\,,
\label{trece}
\eeq
where the flat metric $\eta_{\,\widehat P \,\widehat Q}$ is such that
$\eta_{\,\widehat +\,\widehat -}=1$. The values of these vierbeins are:
\bear
&&E_{-}^{\,\widehat -}\,=\,1\,\,,
\,\,\,\,\,\,\,\,\,\,\,\,\,\,
E_{+}^{\,\widehat -}\,=\,{W\over 2}\,\,,\rc\rc
&&E_{+}^{\,\widehat +}\,=\,1\,\,,
\,\,\,\,\,\,\,\,\,\,\,\,\,\,\,\,\,\,
E_{-}^{\,\widehat +}\,=\,0\,\,,
\,\,\,\,\,\,\,\,\,\,\,\,\,\,\,\,\,\,
E_{i}^{\,\widehat j}\,=\,\delta_{ij}\,\,.
\label{catorce}
\eear

Embedding the M2-brane in such
a way that the worldvolume coordinates are $\xi^{\mu}\,=\,(\,x^{+}, x^{-},
x^{a}\,)$ with the other coordinates being constant, we get:
\beq
\Gamma_{\kappa}^{M2}\,=\,
\Gamma_{\,\widehat -\,\,\widehat +\,\widehat a}\,\,.
\label{quince}
\eeq

Let us next consider an M5-brane probe in the so-called PST formalism
\cite{PST}. We will
take the worldvolume 3-form $H$ of this approach equal to zero. If $a$
is the PST scalar \cite{PST}, the kappa symmetry matrix is:
\beq
\Gamma_{\kappa}^{M5}\,=\,{1\over 5!\,\sqrt{-\det g}}\,
{1\over (\,\partial\cdot a)^2}\,\partial_m\,a\,\gamma^m\,
\gamma_{i_1\cdots i_5}\,\epsilon^{i_1\cdots i_5 n}\,\partial_n\,a\,\,.
\label{dseis}
\eeq
We will embed the M5-brane in such a way that the worldvolume coordinates are
$\xi^{\mu}\,=\,(\,x^{+}, x^{-},x^{a_1},\cdots,x^{a_4}\,)$, with the other
$x^i$'s  constant. The field $a$ can be gauge-fixed to some convenient
value \cite{PST}. Let us take it to be
$a=x^{a_4}$, \ie\ $a$ is equal to the ``last" worldvolume spatial coordinate.
Then, the kappa symmetry matrix (\ref{dseis}) takes the form:
\beq
\Gamma_{\kappa}^{M5}\,=\,
\Gamma_{\,\widehat -\,\,\widehat +\,\widehat a_1\cdots \widehat a_4}\,\,.
\label{dsiete}
\eeq

The M-branes can be extended along only one of the light cone coordinates.
Notice that it cannot be extended only along the $x^-$ coordinate since, as
the $x^-x^-$ component of the metric is zero, the worldvolume metric would be
degenerate (with vanishing determinant) and the corresponding configuration is
not admissible. Therefore, only  M-branes  extended  along
$x^+$ and two other  transverse coordinates are, in principle, possible.
The induced matrix along the light cone coordinate for such a configuration
is:
\beq
\gamma_{+}\,=\,\Gamma_{\,\widehat +}\,+\,
{W\over 2}\,\Gamma_{\,\widehat -}\,\,.
\label{docho}
\eeq
Notice the dependence of $\gamma_{+}$ on the transverse coordinates. This
dependence is transmitted to the kappa symmetry matrix. For example, for  a
M2-brane extended along  the coordinates $(x^+, x^{a_1}, x^{a_2})$, the matrix
$\Gamma_{\kappa}$ is:
\beq
\Gamma_{\kappa}\,=\,{1\over  \sqrt{-W}}\,\,
\Big[\,\Gamma_{\,\widehat +}\,+\,{W\over 2}\,
\Gamma_{\,\widehat -}\,\,\Big]\,\,
\Gamma_{\,\widehat x^{\,a_1} \,\widehat x^{\,a_2}}\,\,.
\label{dnueve}
\eeq
Due to this extra coordinate dependence it is impossible to realize the kappa
symmetry condition $\Gamma_{\kappa}\epsilon=\epsilon$. The same happens for
an M5-brane embedding of this type. Therefore, in what follows,  we would only
consider M2- and M5-branes extended along the two light-cone coordinates. The
corresponding kappa symmetry matrices will be given by the constant matrices
written in eqs. (\ref{quince}) and (\ref{dsiete}) respectively.

Let us rewrite the Killing spinors (\ref{nueve}) as:
\beq
\epsilon\,=\,e^{x^i\Omega_i}\,\chi^{(+)}\,\,,
\label{veinte}
\eeq
where we have taken into account that $\Omega_i\Omega_j=0$ and $\chi^{(+)}$ is
given by:
\beq
\chi^{(+)}\,=\,e^{x^+\Omega_+}\,\chi\,\,.
\label{vuno}
\eeq
In terms of $\chi^{(+)}$ the condition $\Gamma_{\kappa}\epsilon=\epsilon$
can be
written as:
\beq
e^{-x^i\Omega_i}\,\Gamma_{\kappa}\,e^{x^i\Omega_i}\,
\chi^{(+)}\,=\,\chi^{(+)}\,\,.
\label{vdos}
\eeq
Expanding the exponentials on the right-hand side of (\ref{vdos}),  and
comparing
the dependence on the coordinates $x^i$ of both sides of the equation, we get:
\beq
\Gamma_{\kappa}\,\chi^{(+)}\,=\,\chi^{(+)}\,\,,
\,\,\,\,\,\,\,\,\,\,\,\,\,
[\,\Gamma_{\kappa}\,,\,\Omega_i\,]\,\chi^{(+)}\,=\,0\,\,,
\,\,\,\,\,\,\,\,\,\,\,\,\,
\Omega_i\,\Gamma_{\kappa}\,\Omega_i\,\chi^{(+)}\,=\,0\,\,.
\label{vtres}
\eeq
The last condition in (\ref{vtres}) is automatic for the type of embeddings we
are considering, since $\Gamma_{\,\widehat -}^{2}=0$. Let us consider the first
two conditions. Taking $x^{+}=0$ on these equations we get the following
algebraic conditions on the constant spinor $\chi$:
\beq
\Gamma_{\kappa}\,\chi\,=\,\chi\,\,,
\,\,\,\,\,\,\,\,\,\,\,\,\,\,\,\,\,\,\,\,\,\,\,\,\,\,
[\,\Gamma_{\kappa}\,,\,\Omega_i\,]\,\chi\,=\,0\,\,.
\label{vcuatro}
\eeq
Moreover, taking into account that $\Gamma_{\kappa}^2=1$ and the first equation
in (\ref{vcuatro}), one easily proves that
$\Gamma_{\kappa}\,\chi^{(+)}\,=\,\chi^{(+)}$ is equivalent to
\beq
e^{x^+\Gamma_{\kappa}\Omega_+\Gamma_{\kappa}}\,\chi\,=\,
e^{x^+\Omega_+}\,\chi\,\,,
\label{vcinco}
\eeq
which, in turn, is satisfied if and only if:
\beq
[\,\Gamma_{\kappa}\,,\,\Omega_+\,]\,\chi\,=\,0\,\,.
\label{vseis}
\eeq
Similarly, one can prove that
$[\,[\,\Gamma_{\kappa}\,,\,\Omega_i\,]\,,\Omega_+\,]\,\chi\,=\,0$ and,
after taking into account that
$[\,\Gamma_{\kappa}\,,\,\Omega_i\,]\,\chi\,=\,0$,
one concludes that we must have:
\beq
[\,\Gamma_{\kappa}\,,\,\Omega_i\,]\,\Omega_+\,\chi\,=\,0\,\,.
\label{vsiete}
\eeq
The algebraic equations (\ref{vcuatro}), (\ref{vseis}) and (\ref{vsiete}) for
the constant spinor $\chi$ are equivalent to the kappa symmetry condition
$\Gamma_{\kappa}\,\epsilon=\epsilon$ and  will be the starting point of our
analysis of the supersymmetry preserved by the different brane probe
configurations. First of all,  notice that, from the expression of
$\Omega_+$ and
the fact that $\Gamma_{\kappa}$ always commutes with
$\Gamma_{\,\widehat -}\,\Gamma_{\,\widehat +}$, eq.
(\ref{vseis}) can be written as:
\beq
[\,\Gamma_{\kappa}\,,\,\theta\,]\,\big(\,
\Gamma_{\,\widehat -}\,\Gamma_{\,\widehat +}\,+\,1\,)\,\chi\,=\,0\,\,.
\label{vocho}
\eeq
Moreover, the matrix $\Gamma_{\,\widehat -}\,\Gamma_{\,\widehat +}\,+\,1$ has
no non-trivial zero modes. Indeed, if $\chi$ is such a zero mode, it would
satisfy $\Gamma_{\,\widehat -}\,\Gamma_{\,\widehat +}\,\chi=-\chi$. By
multiplying this last equation by $\Gamma_{\,\widehat +}$, and using the
anticommutation relation $\{\Gamma_{\,\widehat +}, \Gamma_{\,\widehat -}\}=2$,
one obtains $\Gamma_{\,\widehat +}\,\chi=0$. Plugging this result in the
zero-mode equation one gets that $\chi=0$. Actually,
the matrix $\Gamma_{\,\widehat -}\,\Gamma_{\,\widehat +}\,+\,1$ is invertible
and its inverse is
$(\Gamma_{\,\widehat +}\,\Gamma_{\,\widehat -}\,+\,1)/3$. It follows that one
must have:
\beq
[\,\Gamma_{\kappa}\,,\,\theta\,]\,\chi\,=\,0\,.
\label{vnueve}
\eeq
Following the same steps we can also prove that eq. (\ref{vsiete}) is
equivalent
to:
\beq
[\,\Gamma_{\kappa}\,,\,\Omega_i\,]\,\,\theta\,\chi\,=\,0\,\,.
\label{treinta}
\eeq
Notice that to satisfy eq. (\ref{vnueve}) either
$\Gamma_{\kappa}$ and $\theta$ commute or else $\chi$ is a zero mode of
$[\,\Gamma_{\kappa}\,,\,\theta\,]$. To study the appearance of such zero modes,
let us split $\theta$ in two pieces, $\theta=\theta\,'+\theta\,''$, such that:
\beq
\{\Gamma_{\kappa}\,,\,\theta\,'\}\,=0\,\,,
\,\,\,\,\,\,\,\,\,\,\,\,\,\,\,\,\,\,\,\,\,\,\,\,\,\,
[\Gamma_{\kappa}\,,\,\theta\,'']\,=0\,\,.
\label{tuno}
\eeq
Then, it is clear that
$[\Gamma_{\kappa}\,,\,\theta\,]\,=-2\theta\,'\Gamma_{\kappa}$ and eq.
(\ref{vnueve}) implies that $\chi$ must be a zero mode of $\theta\,'$.
Let us similarly split the $\Omega_i$'s as
$\Omega_i=\Omega_i'+\Omega_i''$, where $\Omega_i'$ ($\Omega_i''$) is
given by the second expression in (\ref{diez}) with $\theta$ substituted by
$\theta\,'$($\theta\,''$).
In order to study the algebraic conditions involving the $\Omega_i$'s, it is
important to distinguish between coordinates along the worldvolume of the brane
an those orthogonal to it.  Accordingly, let us
split the  $x^i$ 's as $x^i\,=\,(x^a, x^{\alpha})$, where $x^a$ are the
coordinates along which the M-brane is extended and the $x^{\alpha}$'s are
constant and determine the location of the brane in the transverse space. It is
important to point out that, when the brane is placed at $x^{\alpha}=0$, we
should consider only the conditions involving the $\Omega_a$'s. Moreover,
since
$[\,\Gamma_{\kappa}\,,\,\Gamma_{\,\widehat -\,\widehat a}\,]=
\{\,\Gamma_{\kappa}\,,\,\Gamma_{\,\widehat -\,\widehat \alpha}\,\}=0$, one has:
\bear
&&[\Gamma_{\kappa}\,,\,\Omega_{a}'\,]\,=-2\Omega_{a}\,'\,\Gamma_{\kappa}\,\,,
\,\,\,\,\,\,\,\,\,\,\,\,\,\,\,\,\,\,\,\,\,\,\,\,\,\,
[\Gamma_{\kappa}\,,\,\Omega_{a}''\,]\,=\,0\rc\rc
&&[\Gamma_{\kappa}\,,\,\Omega_{a}'\,]\,=\,0\,\,,
\,\,\,\,\,\,\,\,\,\,\,\,\,\,\,\,\,\,\,\,\,\,\,\,\,\,\,\,\,
\,\,\,\,\,\,\,\,\,\,\,\,\,\,\,\,\,\,
[\Gamma_{\kappa}\,,\,\Omega_{\alpha}''\,]\,=
-2\Omega_{\alpha}''\,\Gamma_{\kappa}\,\,.
\label{tdos}
\eear
It is now straightforward to reduce the conditions (\ref{vcuatro}),
(\ref{vnueve}) and (\ref{treinta}) to the following set of equations:

\beq
\fbox{$
\begin{array}{rclcrcl}
\Gamma_{\kappa}\,\chi &=&\chi& ~~~~~& \theta\,'\,\chi&=& 0 \\
\Omega_{a}'\,\,\chi&=&0 &\rule{0mm}{6mm} &\Omega_{a}'\,\theta''\,\chi&=&0
\\
\Omega_{\alpha}''\,\chi&=&0 &\rule{0mm}{6mm}
&\Omega_{\alpha}''\theta''\,\chi &=& 0 .\\
\end{array}
$}
\label{ttres}
\eeq
 
\bigskip

In particular, when $\Gamma_{\kappa}$  commutes (anticommutes) with
$\theta$ (\ie\ when $\theta'$($\theta''$) vanishes) the conditions
involving
$\Omega_a$ ($\Omega_{\alpha})$ are absent and the system (\ref{ttres})
collapses to one of the following two lines, in
addition to the equation
$\Gamma_{\kappa}\chi=\chi$:
\beq
\begin{array}{rcc}
[\,\Gamma_{\kappa}\,,\,\theta\,]\,=\,0 &
\Longrightarrow & \Omega_{\alpha}\,\chi\,=0 \,,
 \,~~~\Omega_{\alpha}\,\theta\,\chi\,=0 \\
\{\,\Gamma_{\kappa}\,,\,\theta\,\}\,=\,0 &\rule{0mm}{8mm}
\Longrightarrow & ~~~~~\theta\,\chi\,=0 ,~~~~~~
\Omega_{a}\,\chi\,=0\,\,.\\
\end{array}
\label{tcuatro}
\eeq
The analysis of eqs. (\ref{ttres}) for the different brane embeddings will
allow us to determine their supersymmetry. Actually, for the pp-wave
backgrounds
studied in the main text, the simplified equations (\ref{tcuatro}) will be
enough and we will be able to identify easily those configurations which
preserve
some amount of supersymmetry.

\subsection{Maximally Supersymmetric pp-Wave}
The metric of the maximally supersymmetric pp-wave background in M-theory is
\cite{KG}:

\beq
ds^2_{11}\,=\,2dx^{+}\,dx^{-}\,-\,\Big[\,\big(\,{\mu\over 3}\big)^2\,\vec
y^{\,2}\,+\,
\big(\,{\mu\over 6}\big)^2\,\vec z^{\,2}\,\Big]\,(\,dx^{+}\,)^2\,+\,
d\vec y^{\,2}\,+\,d\vec z^{\,2}\,\,,
\label{tcinco}
\eeq
where $\mu$ is a scale, $\vec y\,=\,(\,y^1\,,\,y^2\,,\,y^3\,)$ and
$\vec z\,=\,(\,z^1\,,\cdots,\,z^6\,)$. We have labeled the transverse
coordinates $x^{i}$ as
$x^{i}=y^i$ for $i=1,\cdots 3$ and $x^{3+j}=z^j$ for $j=1,\cdots 6$. The
four-form F takes the value:
\beq
F\,=\,\mu\,dx^{+}\wedge dy^1\wedge dy^2\wedge dy^3\,\,.
\label{tseis}
\eeq
This background can be obtained by taking $\mu_1=\mu$ and
$\mu_2=\mu_3=\mu_4=0$ in our four parameter ansatz of eqs.
(\ref{cinco})-(\ref{siete}). Notice that the matrix $\theta$, defined in
eq. (\ref{ocho}), is now given by:
\beq
\theta\,=\,\mu\,\Gamma_{\,\widehat y^{\,1}\,\,\widehat y^{\,2}\,\,
\widehat y^{\,3}}\,\equiv\,\mu\,I\,\,,
\label{tsiete}
\eeq
where we have defined the matrix $I$. Moreover, by using the value of $\theta$
given in eq. (\ref{tsiete}) in the definition of the $\Omega_i$'s
(eq. (\ref{diez})), one easily obtains their expressions, namely:
\beq
\Omega_{y^i}\,=\,{\mu\over 6}\,I\, \Gamma_{\,\widehat y^{\,i}}\,
\Gamma_{\,\widehat -}\,\,,
\,\,\,\,\,\,\,\,\,\,\,\,\,\,
\Omega_{z^j}\,=\,{\mu\over 12}\,I\, \Gamma_{\,\widehat z^{\,j}}\,
\Gamma_{\,\widehat -}\,\,.
\label{tocho}
\eeq
The Killing spinors for this supergravity solution are given by the general
expression (\ref{nueve}), where $\chi$  is an arbitrary constant spinor.
Therefore, it has 32 supersymmetries and, actually, it can be obtained by
performing the Penrose limit of the $AdS_4\times S^7$ and
$AdS_7\times S^4$ solutions \cite{Figueroados}.

The SUSY properties of the brane probes will depend of
the spatial  directions occupied by   the branes on the $3+6$ split.
We shall consider  test  M2 and M5 branes in this background extended along
the directions $+$, $-$, $m$ coordinates $y^a$ and $n$ coordinates $z^b$. We
shall denote these configurations as  $(+,-,m,n)$ branes. Clearly $m+n=1$ for a
M2-brane, whereas $m+n=4$ for a M5-brane. Thus, the configurations to
explore of
the $(+,-,m,n)$ type for the M2-brane are:
\beq
(+,-,1,0)\,,
\,\,\,\,\,\,\,\,\,\,\,\,\,\,\,
(+,-,0,1)\,.
\label{tnueve}
\eeq
For  the M5-brane we have the following possibilities of the $(+,-,m,n)$
type:
\bear
&&
(+,-,3,1)\,,
\,\,\,\,\,\,\,\,\,\,\,\,\,\,\,
(+,-,2,2)\,,\rc\rc
&&(+,-,1,3)\,,
\,\,\,\,\,\,\,\,\,\,\,\,\,\,\,
(+,-,0,4)\,.
\label{cuarenta}
\eear
For such a $(+,-,m,n)$ brane configuration we will take the following set of
worldvolume
coordinates:
\beq
\xi^i\,=\,(\,x^{+},x^{-}, y^{a_1},\cdots,y^{a_m},z^{b_1},\cdots,
z^{b_n}\,)\,\,,
\label{cuno}
\eeq
while the other $y$'s and $z$'s are transverse constant scalars. First of
all we
will consider all possible configurations with all these
scalars  equal to zero and, afterwards,  we shall explore the possibility of
giving them a non vanishing value.

We shall apply in our analysis the  methodology which we have developed for the
general case. First of all, we consider the possible ways of fulfilling eq.
(\ref{vnueve}). A kappa symmetry matrix $\Gamma_{\kappa}$ of the types
written in eqs. (\ref{quince}) and (\ref{dsiete}) either commutes or
anticommutes with the matrix $\theta$ of eq. (\ref{tsiete}). Notice that
$\theta$ has no zero modes, since its eigenvalues are $\pm\mu$. Thus, if
$\{\Gamma_{\kappa},\theta\}=0$ eq. (\ref{vnueve}) has no solution, and the
only possibility left is that $[\,\Gamma_{\kappa},\theta\,]=0$. Then,
according to eq. (\ref{tcuatro}), these configurations without transverse
scalars
are $1/2$ supersymmetric, with 16 supersymmetries, which correspond to the
spinors satisfying $\Gamma_{\kappa}\chi=\chi$.

If the brane probe  is placed at a non-zero value of the transverse coordinates
$y^{\alpha}$ and $z^{\alpha}$, we must study the zero modes of
$\Omega_{y^{\alpha}}$ and $\Omega_{z^{\alpha}}$ (see eq. (\ref{tcuatro})). By
inspecting the form of these matrices in eq. (\ref{tocho}), one readily
realizes that $\chi$ is a zero mode of them if and only if
$\Gamma_{\,\widehat -}\,\chi=0$, which means that the corresponding Killing
spinors are all standard.  Thus, we are led to introduce a second projection on
$\chi$ and, as a consequence, the configuration preserves 8 supersymmetries,
\ie\ is $1/4$ supersymmetric.

At this point it is interesting to notice that, as
$\{\,\Gamma_{\,\widehat +},\Gamma_{\,\widehat -}\,\}\,=\,2$,  any spinor $\chi$
can be decomposed as $\chi=\chi_{+}+\chi_{-}$ with:
\beq
\chi_{\pm}\,=\,{1\over 2}\,
\Gamma_{\,\widehat \pm}\,\Gamma_{\,\widehat \mp}\,\chi\,\equiv\,
{\cal P}_{\pm}\,\chi\,\,.
\label{cdos}
\eeq
The operators
${\cal P}_{\pm}\,\equiv\,{1\over 2}\,
\Gamma_{\,\widehat \pm}\,\Gamma_{\,\widehat \mp}$
are clearly projectors, since ${\cal P}_{+}+{\cal P}_{-}=1$,
${\cal P}_{+}{\cal P}_{-}={\cal P}_{-}{\cal P}_{+}=0$ and
$\big({\cal P}_{\pm}\big)^2={\cal P}_{\pm}$. As
$\Gamma_{\,\widehat  \pm}\,\chi_{\pm}\,=\,0$,
it is clear that ${\cal P}_{\pm}$ projects on the subspace of spinors such that
$\Gamma_{\,\widehat \pm}\,\chi_{\pm}\,=\,0$.
Actually, by multiplying the condition
$\Gamma_{\,\widehat \pm}\,\chi_{\pm}\,=\,0$ by
$\Gamma_{\,\widehat \mp}$ and using that
$\Gamma_{\,\widehat \mp}\Gamma_{\,\widehat \pm}=
-\Gamma_{\,\widehat \pm}\Gamma_{\,\widehat \mp}+2$,
one gets
$\Gamma_{\,\widehat \pm}\Gamma_{\,\widehat \mp}\,\chi_{\pm}\,
=\,2\chi_{\pm}$, \ie\ the condition
$\Gamma_{\,\widehat \pm}\,\chi_{\pm}\,=\,0$ is equivalent to
${\cal P}_{\pm}\chi_{\pm}=\chi_{\pm}$. On the other hand, it is interesting to
notice that the condition ${\cal P}_{\pm}\chi_{\pm}=\chi_{\pm}$ is equivalent
to $\Gamma_{\,\widehat \pm\,\widehat \mp}\,\chi_{\pm}=\chi_{\pm}$.

It follows from the above discussion that when
$[\,\Gamma_{\kappa}\,,\theta]\,=0$ and the brane probe is placed at an
arbitrary point in transverse space, the 8 supersymmetries of the system are
characterized by spinors $\chi$ which satisfy:
\beq
\Gamma_{\kappa}\,\chi\,=\,{\cal P}_{-}\chi=\chi\,\,.
\label{ctres}
\eeq
As a consistency check of the projection (\ref{ctres}), one easily verifies
that $[\Gamma_{\kappa}, {\cal P}_{-}]\,=\,0$.

In conclusion, we have to analyze in each case whether or not $\Gamma_{\kappa}$
commutes with the matrix $\theta$ of eq. (\ref{tsiete}). We will do it
separately for M2 and M5 branes in the next subsections. The same results for
the probe analysis in this maximally supersymmetric pp-wave have been found in
ref. \cite{Kim}.

\subsubsection{M2-brane configurations}
\medskip

Let us study the supersymmetry preserved by a $(+,-,1,0)$ configuration.
Without loss of generality we can assume that the M2-brane is extended
along the
direction $y^1$. The corresponding  $\Gamma_{\kappa}$ matrix is:
\beq\,
\Gamma^{M2}_{\kappa}\,=\,\Gamma_{\,\widehat -\,\widehat+}
\Gamma_{\,\widehat y^{\,1}}\,\,.
\label{ccuatro}
\eeq
It is straightforward to verify that the matrix $\Gamma_{\kappa}$ displayed in
eq. (\ref{ccuatro}) commutes with $I$ and, thus, with $\theta$. Therefore,
according to our general analysis, this configuration preserves 16
supersymmetries when the probe  is located at the origin of the transverse
space
and 8 supersymmetries when the M2-brane is placed at an arbitrary point.

It is also immediate to check that the kappa symmetry matrix of
the remaining $(+,-,0,1)$ M2-brane configuration
does not commute with $\theta$ and, as a consequence,
is non-supersymmetric.

\subsubsection{M5-brane configurations}
\medskip
For a $(+,-,m,n)$ configuration the kappa symmetry matrix for
the M5-brane is:
\beq
\Gamma_{\kappa}^{M5}\,=\,\Gamma_{\,\widehat -\,\widehat+}\,\,
\Gamma_{\,\widehat y^{\,a_1}\cdots \,\widehat y^{\,a_m}}\,
\Gamma_{\,\widehat z^{\,b_1}\cdots \,\widehat z^{\,b_n}}\,.
\label{ccinco}
\eeq

After some calculation one can check  that:
\beq
\Gamma_{\kappa}\,I\,=\,(-1)^n\,I\,\Gamma_{\kappa}\,\,,
\label{cseis}
\eeq
and, thus, $\Gamma_{\kappa}$ commutes with $\theta$ only if $n\in 2\ZZ$ (or if
$m\in 2\ZZ$ since $n+m$ is even). Therefore
we conclude that the configurations with $n$ even and located at the origin of
the transverse space are 1/2 supersymmetric.  They are:
\beq
(+,-,2,2)\,,
\,\,\,\,\,\,\,\,\,\,\,\,\,\,\,
(+,-,0,4)\,,
\label{csiete}
\eeq
while these same embeddings with excited transverse scalars are only
1/4 supersymmetric.

In the following table we summarize our results for M-branes in the maximally
supersymmetric pp-wave background. We include only the configurations
which preserve some amount of supersymmetry.

\bigskip
\beq
\begin{tabular}{|c|c|c|}\hline
~ &  \# Susys & \# Susys  \\
M2 &  without scalars & with scalars
 \\ \hline
$(+,-,1,0)$ &  16 & 8
\rule{0mm}{7mm}\\
\hline
M5 &   &
\rule{0mm}{7mm}\\
\hline
$(+,-,2,2)$ &  16 & 8
\rule{0mm}{7mm}\\
$(+,-,0,4)$ &  16 & 8
\rule{0mm}{7mm}\\
\hline
\end{tabular}
\label{tableextra}
\eeq

\subsection{pp-Wave with 24 supersymmetries}
\medskip
Let us now split the coordinates of the eleven dimensional spacetime as:
$x^{\mu}\,=\,(x^{+}\,,\,x^{-}\,,\,x^{i})\,=\,
(x^{+}\,,\,x^{-}\,,\, \vec y\,,\,\vec z\,,\,x^9)$, where $\vec y$ and $\vec z$
are four component vectors, \ie,  $\vec y\,=\,(y^1,y^2,y^3,y^4)$ and
$\vec z\,=\,(z^1,z^2,z^3,z^4)$. Clearly $x^{i}=y^{i}$ and
$x^{4+i}=z^{i}$ for $i=1,\cdots, 4$. The metric of the pp wave with 24
supersymmetries is:
\beq
ds^2_{11}\,=\,2dx^{+}\,dx^{-}\,-\,{\mu^2\over
4}\,\vec y^{\,2}\,(\,dx^{+}\,)^2\,
+\, d\vec y^{\,2}\,+\,d\vec z^{\,2}\,+\,(dx^{9})^2\,\,,
\label{cocho}
\eeq
and the four-form $F$ is:
\beq
F\,=\,\mu\,\big[\,dx^{+}\wedge dy^{1}\wedge  dy^{2}
\wedge dx^{9}\,+\,
dx^{+}\wedge dy^{3}\wedge  dy^{4}\wedge dx^{9}
\,\big]\,\,.
\label{cnueve}
\eeq
This supergravity solution can be obtained from the general four parameter
ansatz of eqs. (\ref{cinco})-(\ref{siete}) by taking $\mu_1=-\mu_2=\mu$,
$\mu_3=-\mu_4=0$, after a suitable relabeling of the transverse coordinates.
In order to characterize the supersymmetry of this background, let us introduce
the following matrix:
\beq
J\,\equiv\,\Gamma_{\widehat y^{\,1}}\,
\Gamma_{\widehat y^{\,2}}\,\Gamma_{\widehat x^{\,9}}\,+\,
\Gamma_{\widehat y^{\,3}}\,
\Gamma_{\widehat y^{\,4}}\,\Gamma_{\widehat x^{\,9}}\,\,,
\label{cincuenta}
\eeq
in terms of which $\theta$ is simply:
\beq
\theta\,=\,\mu\,J\,\,,
\label{ciuno}
\eeq
and the matrices $\Omega_i$ are given by:
\beq
\Omega_{y^i}\,=\,{\mu\over 24}\,[\,3J\Gamma_{\widehat y^{\,i}}\,+\,
\Gamma_{\widehat y^{\,i}}\,J\,]\,\Gamma_{\widehat -}\,\,,
\,\,\,\,\,\,\,\,\,\,\,\,\,
\Omega_{z^j}\,=\,{\mu\over 12}\,J\,\Gamma_{\widehat z^{\,j}}\,
\Gamma_{\widehat -}\,\,,
\,\,\,\,\,\,\,\,\,\,\,\,\,
\Omega_{x^9}\,=\,{\mu\over 6}\,J\,\Gamma_{\widehat x^{\,9}}\,
\Gamma_{\widehat -}\,\,,
\label{cidos}
\eeq

The standard Killing spinors are, in this case,  16 spinors of the form:
\beq
\epsilon^{st}\,=\,e^{-{\mu\over 4}\,x^{+}\,J}\,\,\chi^{st}\,\,,
\,\,\,\,\,\,\,\,\,\,\,\,\,\,\,\,\,\,\,\,\,\,\,\,\,\,
\Gamma_{\widehat -}\,\chi^{st}\,=\,0
\label{citres}
\eeq
This background has, in addition,  8
supernumerary Killing spinors. They are of the form:
\beq
\epsilon^{sn}\,=\,\big(\,1\,+\,{\mu\over 8}\,\Gamma_{-}\,J\,
\sum_{i=1}^{4}\,y^i\,\Gamma_{y^i}\,\big)\,\chi^{sn}
\label{cicuatro}
\eeq
where $\chi^{sn}$ is a constant spinor such that
$\Gamma_{\widehat -}\,\chi^{sn}\,\not= 0$ and which satisfies the condition:
\beq
J\,\chi^{sn}\,=\,0\,\,.
\label{cicinco}
\eeq
Notice that the
supernumerary spinors are all independent of $x^+$. On the contrary the
standard Killing spinors depend on $x^+$ except when $J\chi^{sn} =0$.
Thus, in this case we have 8 standard Killing spinors independent of $x^+$.
Moreover, if we define the matrix $\Gamma^{(y)}$ as:
\beq
\Gamma^{(y)}\equiv
\Gamma_{\widehat y^{\,1}\,\widehat y^{\,2}\,
\widehat y^{\,3}\,\widehat y^{\,4}}\,\,,
\label{cisiete}
\eeq
one can immediately prove that:
\beq
J\chi=0\,\,\,\,\,\,\,\,\,\,\,\,\,\,\,\,\,
\Leftrightarrow
\,\,\,\,\,\,\,\,\,\,\,\,\,\,\,\,\,
\Gamma^{(y)}\,\,\chi\,=\,\chi\,\,.
\label{ciocho}
\eeq

Let us consider an M-brane probe in the previous background. Notice that the
9 transverse dimensions are split as 4+4+1. Actually, the  four $\vec
y$ coordinates are not equivalent and, in general, it is important to
distinguish between the two sets of $y^i$ coordinates. We shall denote by
$(+,-, (m_1,m_2), n, p)$ to a M-brane embedding along $m_1$ coordinates of the
set $\{y^1,y^2\}$ and $m_2$ coordinates of the set $\{y^3,y^4\}$, with  $n$ and
$p$ being the number of $z^i$ and $x^9$ coordinates respectively. Obviously
$m_1+m_2+n+p=1$ for an M2-brane and $m_1+m_2+n+p=4$ for an M5-brane. Moreover,
we have the following equivalence relation:
\beq
(+,-, (m_1,m_2), n, p)\,\approx\,
(+,-, (m_2,m_1), n, p)\,\,.
\label{ciseis}
\eeq

In order to study the number of supersymmetries of the background preserved by
the probe, let us come back to our general formalism and, in particular, to
eq.
(\ref{vnueve}). An important remark concerning this equation is that now
$\theta$ has zero modes. Indeed $\theta\chi=0$ iff $J\chi=0$.
Thus,  according to eq. (\ref{ciocho}), it is possible to solve eq.
(\ref{vnueve}) when
$\{\Gamma_{\kappa}\,,\,\theta\}=0$, provided we require that
$\Gamma^{(y)}\,\chi=\chi$. On the contrary, if $\Gamma_{\kappa}$ commutes with
one term in $J$ and anticommutes with the other, it is impossible to find a
spinor $\chi$ satisfying (\ref{vnueve}). Indeed, in this case
$[\Gamma_{\kappa}\,,\,\theta]=-2\theta'\Gamma_{\kappa}$ reduces to a single
antisymmetrized product of $\Gamma$-matrices, which has no zero modes.

We are thus led to consider  the two possible situations of eq.
(\ref{tcuatro}). Notice that, when $[\Gamma_{\kappa}\,,\,\theta]=0$ and the
brane is located at the origin of coordinates, the only additional condition
required to
$\chi$ is $\Gamma_{\kappa}\,\chi=\chi$ and, therefore, the number of
supersymmetries in this case is 1/2 of that of the background, \ie\ 12. On the
other hand, if $\{\Gamma_{\kappa}\,,\,\theta\}=0$, the conditions in
(\ref{tcuatro}) do not involve the $\Omega_{\alpha}$'s and, thus,
the number of supersymmetries does not change when we move  the brane away from
the origin.

Another interesting observation to study the fulfillment of eq.
(\ref{tcuatro}) is the fact that the matrices $\Omega_{y^i}$ cannot have
supernumerary zero modes. Indeed, a supernumerary zero mode $\chi^{sn}$ of,
say, $\Omega_{y^1}$ must be a zero mode of
$3J\Gamma_{\,\widehat y^{\, 1}}\,+\,\Gamma_{\,\widehat y^{\, 1}}J$. After
taking into account the explicit expression of $J$ (eq. (\ref{cincuenta})), one
immediately realizes that such a $\chi^{sn}$ must also be a zero mode of
$2\Gamma_{\,\widehat y^{\, 2}}-
\Gamma_{\,\widehat y^{\, 1}}\Gamma_{\,\widehat y^{\, 3}}
\Gamma_{\,\widehat y^{\, 4}}$, which is impossible. Notice that this argument
does not apply to the matrices $\Omega_{z^j}$ and $\Omega_{x^9}$, whose
supernumerary zero modes must satisfy the condition written in eq.
(\ref{ciocho}).

As in the previously studied case, we shall analyze separately the M2 and M5
cases.

\subsubsection{M2-brane configurations}
\medskip
Although the  four $\vec y$ coordinates are not equivalent,
since there is only one transverse coordinate along the M2-brane
worldvolume, and
due to the equivalence (\ref{ciseis}), the distinction among the
$\vec y$ coordinates  is irrelevant in
this M2 case. Accordingly,  we shall denote by $(+,-,m,n,p)$ to a
M2-brane configuration extended along $m$ coordinates
$y$, $n$ coordinates $z$ and $p$ coordinates $x^9$ ($m,n,p=0,1$). It is
obvious that we have the following possibilities:
\beq
(+,-,1,0,0)\,\,,
\,\,\,\,\,\,\,\,\,\,\,\,\,
(+,-,0,1,0)\,\,,
\,\,\,\,\,\,\,\,\,\,\,\,\,
(+,-,0,0,1)\,\,.
\label{cinueve}
\eeq

The kappa symmetry matrix $\Gamma_{\kappa}^{M2}$ was written in general in eq.
(\ref{quince}). A simple calculation yields the result that
$\Gamma_{\kappa}^{M2}$  and $J$ only commute for the $(+,-,0,0,1)$
configuration,  whereas they anticommute
for the $(+,-,0,1,0)$ embedding. Finally, in the $(+,-,1,0,0)$ case,
$\Gamma_{\kappa}^{M2}$ commutes with  one term in $J$  and anticommutes
with the
other and, therefore, this embedding does not preserve any supersymmetry.

Let us first consider in detail the $(+,-,0,1,0)$ system. According to eq.
(\ref{tcuatro}), the spinor $\chi$ must satisfy that $J\chi=0$ and
$\Omega_{z^a}\chi=0$. For standard spinors the second condition is automatic,
while the first one requires the introduction of a new projection. Thus, the
preserved standard spinors $\chi^{st}$ are four constant spinors satisfying the
conditions ${\cal P}_{-}\,\chi^{st}\,=\,\Gamma_{\kappa}^{M2}\,\chi^{st}\,=\,
\Gamma^{(y)}\,\chi^{st}\,=\,\chi^{st}$. Notice that these three projections
commute among themselves, as it should. Moreover, from the form of
$\Omega_{z^a}$ as given in eq. (\ref{cidos}), we learn that the conditions
of eq.
(\ref{tcuatro}) are already satisfied by the supernumerary spinors of the
background and we only have  to impose the condition
$\Gamma_{\kappa}^{M2}\,\chi^{sn}\,=\,\chi^{sn}$, which gives four spinors of
this type. In general, we will say that a configuration is $A(B+C)$
supersymmetric if it preserves $A$ supersymmetries, being $B$($C$) the
number of
them corresponding to standard (supernumerary) spinors (obviously $A=B+C$).
With
this notation the $(+,-,0,1,0)$ system is $8(4+4)$ supersymmetric. Notice that,
according to eq. (\ref{tcuatro}), this configuration preserves the same number
of supersymmetries at any point in transverse space.

Let us now consider the $(+,-,0,0,1)$ configuration. As $\Gamma_{\kappa}^{M2}$
commutes with $\theta$ in this case, this embedding is  $12(8+4)$
supersymmetric
when the brane is located at the origin. The corresponding spinors are the
original ones is eqs. (\ref{citres}) and (\ref{cicuatro}) with the extra
projections  $\Gamma_{\kappa}^{M2}\chi^{st}\,=\,\chi^{st}$ and
$\Gamma_{\kappa}^{M2}\chi^{sn}\,=\,\chi^{sn}$. If constant scalars are excited
we have to impose the condition
$\Omega_{y^{\alpha}}\chi\,=\,0$, which is impossible to satisfy for
supernumerary spinors. Therefore, the supernumerary spinors are lost away from
the origin and we are left with a $8(8+0)$ supersymmetric system\footnote{We
could move the brane away from the origin in the $z$-directions only. In this
case the four supernumerary supersymmetries are still preserved. However,
in what
follows for a configuration with scalars excited we will mean the case in which
the brane is located at a generic point in transverse space.}.

The situation for M2-branes is summarized in the following
table:
\bigskip
\beq
\begin{tabular}{|c|c|c|}\hline
~ &  \# SUSYs & \# SUSYs  \\
Configuration &  without scalars & with scalars
 \\ \hline
$(+,-,0,1,0)$ &  8\,(4+4) & 8\,(4+4)
\rule{0mm}{7mm}\\
\hline
$(+,-,0,0,1)$ &  12\,(8+4) & 8\,(8+0)
\rule{0mm}{7mm}\\
\hline
\end{tabular}
\label{sesenta}
\eeq

We have only included the configurations with some supersymmetry and we have
explicitly indicated the number of standard and supernumerary supersymmetries.

\subsubsection{M5-brane configurations}
\medskip
As in eq. (\ref{dsiete}), the kappa symmetry matrix
$\Gamma_{\kappa}^{M5}$ will be taken as the antisymmetrized product of the
Dirac
matrices along the worldvolume directions.
In order to analyze the supercharges associated to standard Killing
spinors which are preserved, we have to characterize those configurations for
which  $\Gamma_{\kappa}^{M5}$ commutes or anticommutes with $J$. It can be
proved that $[\Gamma_{\kappa}^{M5}, J]\,=\,0$ for:
\bear
&&(+,-, (2,2), 0, 0)\,\,,\rc
&&(+,-, (2,0), 2, 0)\,\approx\,(+,-, (0,2), 2, 0)\,\,,\rc
&&(+,-, (0,0), 4, 0)\,\,,\rc
&&(+,-, (1,1), 1, 1)\,\,.
\label{suno}
\eear
Moreover, $\{\Gamma_{\kappa}^{M5}, J\}\,=\,0$ for:
\bear
&&(+,-, (1,1), 2, 0)\,\,,\rc
&&(+,-, (2,0), 1, 1)\,\approx\,(+,-, (0,2), 1, 1)\,\,,\rc
&&(+,-, (0,0), 3, 1)\,\,.
\label{sdos}
\eear
Notice $[\Gamma_{\kappa}, \Gamma^{(y)}]=0$ if the number of $y^i$ coordinates
in the worldvolume is even. This condition, which   holds for all the
configurations
written above in  (\ref{suno}) and (\ref{sdos}), is needed to ensure the
compatibility between the kappa symmetry projection and the one corresponding
to supernumerary spinors.

The configurations (\ref{suno}) preserve $12(8+4)$ supersymmetries when placed
at the origin $(\tilde x^{\,i}=0)$. Outside the origin they generically
lose the
supersymmetries associated to $\chi^{sn}$, since when moving in the
$y$-directions one is forced to impose that $\Omega_{y}\chi^{sn}=0$. The
exception to this behavior is the $(+,-, (2,2), 0, 0)$ configuration, because
it has no external
$y$-directions (of course if we displace the M5 along an external $z$ or $x^9$
directions these configurations are still $12(8+4)$ supersymmetric).

The supersymmetry of the embeddings listed in (\ref{sdos}) is not changed by
translations in the transverse space. They all have four standard spinors which
correspond to the projections
$\Gamma_{\kappa}\chi^{st}={\cal P}_{-}\chi^{st}=
\Gamma^{(y)}\chi^{st}=\chi^{st}$ and only one of them, namely
$(+,-, (0,0), 3, 1)$, has four supernumerary supersymmetries due to the fact
that it has no worldvolume directions along the $y$ coordinates.

The result of this analysis is summarized in the following table:
\bigskip
\beq
\begin{tabular}{|c|c|c|}\hline
~ &  \# SUSYs & \# SUSYs  \\
Configuration &  without scalars & with scalars
 \\ \hline
$(+,-, (2,2), 0, 0)$ &  12\,(8+4) & 12\,(8+4)
\rule{0mm}{7mm}\\
\hline
$(+,-, (2,0), 2, 0)$ &  12\,(8+4) & 8\,(8+0)
\rule{0mm}{7mm}\\
\hline
$(+,-, (0,0), 4, 0)$ &  12\,(8+4) & 8\,(8+0)
\rule{0mm}{7mm}\\
\hline
$(+,-, (1,1), 1, 1)$ &  12\,(8+4) & 8\,(8+0)
\rule{0mm}{7mm}\\
\hline
$(+,-, (1,1), 2, 0)$ &  4\,(4+0) & 4\,(4+0)
\rule{0mm}{7mm}\\
\hline
$(+,-, (2,0), 1, 1)$ &  4\,(4+0) & 4\,(4+0)
\rule{0mm}{7mm}\\
\hline
$(+,-, (0,0), 3, 1)$ &  8\,(4+4) & 8\,(4+4)
\rule{0mm}{7mm}\\
\hline
\end{tabular}
\label{stres}
\eeq
Again, we have only included in the table the configurations which preserve
some
supersymmetries.

\setcounter{equation}{0}
\section{Supergravity solutions}
\medskip

In this section we are going to develop the formalism needed to obtain
supergravity backgrounds for the brane-wave intersections studied in the
brane probe approach of the previous section. We will also analyze the
degrees of supersymmetry of the different solutions and  compare them
with the ones corresponding to the brane probe.

The eleven dimensional metrics of the solutions we will be dealing with are
warped generalizations of the line element written in eq. (\ref{uno}), namely:
\beq
ds^2_{11}\,=\,h_1\,(\,2dx^+dx^-\,+\,W\,(dx^+)^2\,+(dx^a)^2\,)\,+\,
h_2\,(\,d\tilde x^{\,\alpha}\,)^2\,\,,
\label{scuatro}
\eeq
where we have distinguished between transverse coordinates parallel to the
brane worldvolume ($x^a$) and those orthogonal to it ($\tilde x^{\,\alpha}$).
The warp factors $h_1$ and $h_2$ are taken to depend on the external
coordinates
$\tilde x^{\,\alpha}$, whereas the profile $W$ can depend on both sets of
coordinates. Actually, we will adopt the ansatz in which $h_1$ and $h_2$
are powers of the same function $H$. These powers are different for the M2
and M5 branes:
\bear
&&h_1=H^{-{2\over 3}},\,\,\,\,\,\,\,\,\,\,\,\,\,\,\,\,
h_2=H^{{1\over 3}}\,\,,\,\,\,\,\,\,\,\,({\rm M2})\,,\rc
&&h_1=H^{-{1\over 3}},\,\,\,\,\,\,\,\,\,\,\,\,\,\,\,\,
h_2=H^{{2\over 3}}\,\,,\,\,\,\,\,\,\,\,({\rm M5})\,.
\label{scinco}
\eear
Notice that the warp factors in (\ref{scinco}) are exactly the same ones which
appear in the pure M2 or M5 solutions.
The four-form field $F$ will be taken as a sum of a wave contribution
$F_{wave}$
and a brane contribution .
We shall assume that $F_{wave}$ is given by the same expression as in eq.
(\ref{dos}), \ie:
\beq
F_{wave}\,=\,dx^{+}\wedge \Theta\,\,,
\label{sseis}
\eeq
where $\Theta$ is defined in eq. (\ref{tres}).
For an M2-brane extended along $(x^{+}, x^{-}, x^{a})$, the contribution to the
four-form field strength will be given by the standard ``electric" ansatz:
\beq
F_{M2}\,=\,dx^{+}\wedge dx^{-}\wedge dx^{a}\wedge dH^{-1}\,\,,
\label{ssiete}
\eeq
while for the M5-brane we will adopt the following magnetic ansatz:
\beq
F_{M5}\,=\,{}^{\widetilde *}dH\,\,.
\label{socho}
\eeq
In eq. (\ref{socho}) ${}^{\widetilde *}$ denotes the Hodge dual with respect to
the external coordinates $\tilde x^{\,\alpha}$ with the Euclidean metric.
The total field strength $F$ must satisfy the Bianchi identity, $dF=0$, and
the field equation:
\beq
d\,{}^*F\,=\,{1\over 2}\,F\wedge F\,\,,
\label{snueve}
\eeq
where $*$ denotes the Hodge dual for the eleven dimensional metric
(\ref{scuatro}). These equations for $F$ are enough to fix the precise
dependence of $H$ on the external coordinates $\tilde x^{\,\alpha}$. Indeed, we
will show that, in general, $H$ will only depend on some subset of
the $\tilde x^{\,\alpha}$'s and it will be a harmonic function on the other
external coordinates. The independence of $H$ on some
$\tilde x^{\,\alpha}$'s means that the brane is smeared along those directions
and, thus, our M-branes are delocalised objects in transverse space. This
remark will be relevant when comparing the number of supersymmetries of the
supergravity solutions with those obtained within the brane probe approach.

The metric and gauge field must also satisfy Einstein's equations, which,
written in flat coordinates, read:
\beq
R_{\widehat P\,\widehat Q}\,=\,
{1\over 12}\,
F_{\widehat P}^{\,\,\,\widehat N_1\cdots \widehat N_3}\,\,
F_{\widehat Q \,\,\widehat N_1\cdots \widehat N_3}\,-\,
{1\over 144}\,\,\eta_{\widehat P\,\widehat Q}\,\,F^2\,\,,
\label{setenta}
\eeq
where $R_{\widehat P\,\widehat Q}$ is the Ricci tensor. The components of
this tensor for a metric of the type (\ref{scuatro}) are written in
appendix A. By inspecting eqs. (\ref{apaocho}) and (\ref{apanueve}), one
 realizes that the profile $W$ only enters the
$\widehat +\,\widehat +$ component of the Ricci tensor. Moreover, one easily
concludes that the only contribution to the right-hand side of eq.
(\ref{setenta}) for
$P=Q=+$ comes from $F_{wave}$. With the purpose of writing this
$\widehat +\,\widehat +$ Einstein equation in a simpler form, let us choose a
basis of one-forms $e^{\widehat M}$ as in eq. (\ref{apados}) and let us
introduce the inverse vierbeins $E_{\,\,\widehat P}^{\,M}$ by means of the
relation $dx^M\,=\,E_{\,\,\widehat P}^{\,M}\,\,e^{\widehat P}$. Then, we can
write  $\Theta$ as:
\beq
\Theta\,=\,{1\over 6}\,
\theta_{\,\widehat i \,\,\,\widehat j\, \,\,\widehat k}\,\,
e^{\widehat i}\wedge e^{\widehat j}\wedge e^{\widehat k}\,\,,
\label{stuno}
\eeq
where:
\beq
\theta_{\,\widehat i \,\,\,\widehat j\, \,\,\widehat k}\,=\,
E_{\,\,\widehat i}^{\,l}\,E_{\,\,\widehat j}^{\,m}\,E_{\,\,\widehat k}^{\,n}\,
\,\theta_{lmn}\,\,.
\label{stdos}
\eeq
Then, it is straightforward to show that the
$\widehat +\,\widehat +$ Einstein equation is equivalent to the following
differential equation for the profile $W$:
\beq
\partial_a^2\,W\,+\,H^{-1}\,\partial^2_{\alpha}\,W\,=\,
-{1\over 6}\,
\theta_{\,\widehat i \,\,\,\widehat j\, \,\,\widehat k}\,
\theta^{\,\widehat i \,\,\,\widehat j\, \,\,\widehat k}\,\,.
\label{sttres}
\eeq
(Compare eqs. (\ref{sttres}) and (\ref{cuatro})).

Once $H$ is determined from the equation of motion of the gauge field, eq.
(\ref{sttres}) allows to obtain the profile $W$. Notice that in the passage
from curved to flat components in (\ref{stdos}), new powers of $H$ are
introduced and, thus, the right-hand side of (\ref{sttres}) does, in general,
depend on the $\tilde x^{\,\alpha}$ coordinates. Moreover, it can be verified
that the other components of Einstein's equations are satisfied by our
ansatz of the metric and $F$, provided $H$ is harmonic, both for the M2
and M5 cases. Therefore, the only non-trivial information we get from
(\ref{setenta}) is just the profile equation (\ref{sttres}). The solutions
of this equation are, in general, different from the values of $W$ for the
pure wave. This fact is a manifestation of the back-reaction  exerted on
the profile of the wave by the presence of the brane \cite{Bain}.

Let us analyze the behavior of our solutions under supersymmetry. A bosonic
configuration of eleven dimensional supergravity is invariant under all
supersymmetry transformations which do not change the gravitino. The parameter
of such  transformation is a spinor $\eta$ which satisfies the so-called
Killing spinor equation:
\beq
\nabla_{M}\,\eta\,=\,\Omega_M\,\eta\,\,,
\label{stcuatro}
\eeq
where $\nabla_{M}$ is the covariant derivative and $\Omega_M$ is given by:
\beq
\Omega_M\,=\,{1\over 288}\,F_{PQRS}\,\Big(\,\Gamma^{PQRS}_{\quad \quad M}
\,+\,8\,\Gamma^{PQR}\,\delta_{M}^{S}\,\Big)\,\,.
\label{stcinco}
\eeq
Notice that $\Omega_M$ is linear in the gauge field. Thus, we can split it as:
\beq
\Omega_M\,=\,\Omega_M^{w}\,+\,\Omega_M^{br}\,,
\label{stseis}
\eeq
where $\Omega_M^{w}$ and $\Omega_M^{br}$ are, respectively, the contributions
to $\Omega_M$ of the wave and brane terms of $F$. Let us define $\theta$ as in
eq. (\ref{ocho}) (notice that now $\theta$ can depend on the coordinates
$\widetilde x^{\, \alpha}$). Then, it is straightforward to show that, both for
an M2 or M5 metric, the different components of $\Omega_M^{w}$ are:
\bear
\Omega_{-}^{w}&=&0\,\,,\rc\rc
\Omega_{+}^{w}&=&-{1\over 12}\,\theta\,\Big[\,
\Gamma_{\,\widehat -}\,\Gamma_{\,\widehat +}\,+\,1\,\Big]\,,\rc\rc
\Omega_{a}^{w}&=&{1\over 24}\,\Big[\,3\theta\Gamma_{\,\widehat a}\,+\,
\Gamma_{\,\widehat a}\,\theta\,\Big]\,\Gamma_{\,\widehat -}\,\,,\rc\rc
\Omega_{\alpha}^{w}&=&{H^{{1\over 2}}\over 24}\,
\Big[\,3\theta\Gamma_{\,\widehat \alpha}\,+\,
\Gamma_{\,\widehat \alpha}\,\theta\,\Big]\,\Gamma_{\,\widehat -}\,\,.
\label{stsiete}
\eear
Moreover, it is rather convenient to define a new spinor $\epsilon$, which is
related to $\eta$ by means of the expression:
\beq
\eta\,=\,H^{\Delta}\,\epsilon\,\,,
\label{stocho}
\eeq
where the exponent $\Delta$ is:
\beq
\Delta\,=\,\cases{
-{1\over 6}\,,&for a M2-brane\,\,,\cr\cr
-{1\over 12}\,,&for a M5-brane\,\,.}
\label{stnueve}
\eeq
Let us now plug the ansatz (\ref{stocho}) in the Killing spinor equation
(\ref{stcuatro}). To compute $\Omega_M^{br}$ we use the brane term of the gauge
field,  written in eqs. (\ref{ssiete}) and (\ref{socho}). Moreover, we shall
impose to $\eta$ the corresponding M-brane projection
$\Gamma_{\kappa}\,\eta\,=\, \eta$, where $\Gamma_{\kappa}$ is given in eq.
(\ref{quince}) or (\ref{dsiete}). Then, one can verify that $\Omega_M^{br}$
drops out and we are left with the following set of differential equations
for $\epsilon$:
\bear
\partial_-\,\epsilon&=&\,0\,,\rc\rc
\partial_+\,\epsilon&=&\Big[\,{1\over 4}\,\partial_a W\,
\Gamma_{\,\widehat a\,\,\widehat -}\,+\,
{1\over 4}\,H^{-{1\over 2}}\partial_{\alpha} W\,
\Gamma_{\,\widehat \alpha\,\,\widehat -}\,+\,
\Omega_{+}^{w}\,\Big]\,\epsilon\,,\rc\rc
\partial_a\,\epsilon&=&\Omega_{a}^{w}\,\epsilon\,,\rc
\partial_{\alpha}\,\epsilon&=&\Omega_{\alpha}^{w}\,\epsilon\,\,,
\label{ochenta}
\eear
which, together with the algebraic condition
$\Gamma_{\kappa}\,\epsilon\,=\, \epsilon$, determine the Killing spinor
$\epsilon$. Let us first find the solutions of the system (\ref{ochenta}) which
correspond to standard spinors $\epsilon^{st}$ satisfying the condition
$\Gamma_{\widehat -}\,\epsilon^{st}\,=\,0$. In this case, the previous
equations reduce to:
\bear
&&\partial_-\,\epsilon^{st}\,=\,\partial_a\,\epsilon^{st}\,=\,
\partial_{\alpha}\,\epsilon^{st}\,=\,0\,,\rc\rc
&&\partial_+\,\epsilon^{st}\,=\,-{\theta\over 4}\,\epsilon^{st}\,\,,\rc\rc
&&\Gamma_{\kappa}\,\epsilon^{st}\,=\,\epsilon^{st}\,\,,
\label{ouno}
\eear
from which we get the following integrability conditions:
\bear
[\,\Gamma_{\kappa}\,,\,\theta\,]\,\epsilon^{st}\,&=&0\,\,,\rc\rc
\partial_{\alpha}\,\theta\,\epsilon^{st}&=&0\,\,.
\label{odos}
\eear
If these conditions hold, one finds eight standard spinors of the form:
\beq
\epsilon^{st}\,=\,e^{-x^+\,\,{\theta \over 4}}\,\chi^{st}\,\,,
\,\,\,\,\,\,\,\,\,\,\,\,
\Gamma_{\kappa}\,\chi^{st}\,=\,\chi^{st}\,\,,
\,\,\,\,\,\,\,\,\,\,\,\,
\Gamma_{\widehat -}\,\chi^{st}\,=\,0\,\,,
\label{otres}
\eeq
where $\chi^{st}$ is a constant spinor.

For general Killing spinors it is easy to find the compatibility conditions
between the equations in (\ref{ochenta}) and the condition
$\Gamma_{\kappa}\,\epsilon\,=\,{\epsilon}$. These conditions are:
\bear
&&[\,\Gamma_{\kappa}, \Omega_{a}^{w}\,]\,\epsilon\,=\,
[\,\Gamma_{\kappa}, \Omega_{\alpha}^{w}\,]\,\epsilon\,=\,0\,\,,\rc\rc
&&[\,\Gamma_{\kappa}, \Omega_{+}^{w}\,]\,\epsilon\,=\,
{1\over 2}\,H^{-{1\over 2}}\,\partial_{\alpha}\,W\,
\Gamma_{\,\widehat \alpha\,\,\widehat -}\,\,\epsilon\,\,,
\label{ocuatro}
\eear
where we have taken into account that $\Gamma_{\kappa}$ always commutes
with $\Gamma_{\,\widehat a\,\,\widehat -}$ and anticommutes with
$\Gamma_{\,\widehat \alpha\,\,\widehat -}$.
Other interesting consistency conditions come from the mutual consistency of
eqs. (\ref{ochenta}):
\beq
\partial_{\alpha}\,\Omega_{a}^{w}\,\epsilon\,=\,0\,\,.
\label{ocinco}
\eeq
Eqs. (\ref{ocuatro}) and (\ref{ocinco}) are very useful to discard the
possibility of having Killing spinors is some cases. Indeed, let us assume that
$\Gamma_{\kappa}$ either commutes or anticommutes with $\theta$, as happened in
all the cases studied in the previous section. If
$[\Gamma_{\kappa}, \theta]=0$ it follows that
$[\Gamma_{\kappa}, \Omega_{a}]=\{\Gamma_{\kappa}, \Omega_{a}\}=0$, whereas
when
$\{\Gamma_{\kappa}, \theta\}=0$ one has
$\{\Gamma_{\kappa}, \Omega_{a}\}=[\Gamma_{\kappa}, \Omega_{\alpha}]=0$. By
using these results in the first equation in (\ref{ocuatro}) and in
(\ref{ochenta}), one finds:
\bear
&&[\,\Gamma_{\kappa}\,,\,\theta\,]\,=\,0\,\,\,
\Longrightarrow\,\,\,
\Omega_{\alpha}^{w}\,\epsilon\,=\,0\,\,\,\,
\Longrightarrow\,\,\,
\partial_{\alpha}\,\epsilon\,=\,0\,\,,\rc\rc
&&\{\,\Gamma_{\kappa}\,,\,\theta\,\}\,=\,0\,\,\,
\Longrightarrow\,\,\,
\Omega_{a}^{w}\,\epsilon\,=\,0
\,\,\,\,
\Longrightarrow\,\,\,
\partial_{a}\,\epsilon\,=\,0\,\,,
\label{oseis}
\eear
which means that some $\Omega_i$'s must have a zero mode. This, in some cases,
is impossible, which allows to discard the existence of certain classes of
Killing spinors. Actually, these conditions, although they are not
complete, are
restrictive enough, as we will see in the particular examples studied in
the next
section. Only for those configurations which succeed in passing the test of
eqs.
(\ref{ocuatro})-(\ref{oseis}) we will try to integrate directly the system
(\ref{ochenta}) and, in these cases, a separation of variables is possible and
the solution of (\ref{ochenta}) is easily found.

\setcounter{equation}{0}
\section{M-branes in the maximally SUSY pp-wave}
\medskip

Let us particularize the general formalism of the previous section to the
intersection of M-branes and the maximally supersymmetric pp-wave. In this case
the four-form field strength will be of the form:
\beq
F\,=\,\mu\,dx^+\wedge dy^1\wedge dy^2\wedge dy^3\,+\,
{\rm brane \,\, term}\,\,,
\label{osiete}
\eeq
with $\mu$ being a constant. Let us consider a $(+,-,m,n)$ configuration and
let us  take the worldvolume coordinates to be
$y^{a}\,=\,(\,y^1,\cdots,y^m\,)$ and $z^{a}\,=\,(\,z^1,\cdots,z^n\,)$. The
external coordinates will be $\tilde y^{\,\alpha}=y^{m+\alpha}$ for
$\alpha=1,\cdots ,3-m$ and $\tilde z^{\,\alpha}=z^{n+\alpha}$ for
$\alpha=1,\cdots ,6-n$. The metric will be:
\bear
ds^2_{11}&=&h_1\,\Big(\,2dx^+dx^-\,+\,
W\,(dx^+)^2\,+(dy^a)^2\,+(dz^a)^2\,\Big)\,+\,\rc\rc
&&+\,h_2\,\Big(\,(d\tilde y^{\,\alpha})^2\,+\,(d\tilde z^{\,\alpha})^2\,\Big)
\,\,,
\label{oocho}
\eear
where $h_1$ and $h_2$ are taken as in eq. (\ref{scinco}) in terms of a function
$H$ which must be determined from the gauge field equations. Recall that the
profile $W$ can be obtained by integrating eq. (\ref{sttres}).

We are only interested in solutions which are invariant under some amount of
supersymmetry. It is not difficult to characterize these solutions.
First of all, notice for this pp-wave the matrix $\theta$ is proportional
to the
matrix $I$ defined in eq. (\ref{tsiete}) and, similarly, the $\Omega_i$'s are
proportional to the matrices written in eq. (\ref{tocho}). Since
$\Gamma_{\kappa}$ either commutes or anticommutes with $I$, we are in one of
the situations considered in eq. (\ref{oseis}) and, thus, the Killing spinors
$\epsilon$  must be a zero mode of some of the $\Omega_i$'s. This is not
possible for supernumerary spinors and, thus, we conclude that our solutions
can only have standard spinors. However, according to the first equation in
(\ref{odos}), the latter can only exist if $[\Gamma_{\kappa}, I]=0$, since
otherwise $I\,\chi^{st}=0$, which cannot be satisfied for $\chi^{st}\not=0$.
Thus, we can restrict ourselves to those configurations for which
$\Gamma_{\kappa}$ commutes with $I$, which were precisely the ones studied in
section 2.1. Notice, however, that this condition is not enough, since the
second equation in (\ref{odos}) implies that $\partial_{\alpha}\theta=0$. If
these conditions hold, the corresponding solution will have eight
standard supersymmetries.

\subsection{M2-branes}
\medskip

For a $(+,-,m,n)$ configuration of a M2-brane $(m+n=1)$, it is straightforward
to compute the matrix $\theta$. One gets:
\beq
\theta\,=\,\mu\,H^{{m-1\over 2}}\,
\Gamma_{\,\widehat y^{\,1}\,\widehat y^{\,2}\,\widehat y^{\,3}}\,\,,
\label{onueve}
\eeq
from which we obtain the following equation for
the profile $W$:
\beq
\partial_a^2\,W\,+\,H^{-1}\,\partial_{\alpha}^2\,W\,=\,-\mu^2\,H^{m-1}\,\,.
\label{noventa}
\eeq

\subsubsection{ $(+,-,1,0)$ }
\medskip
We saw in section 2.1 that $\Gamma_{\kappa}^{M2}$
commutes with $I$ only in this $m=1$ case. The complete expression of
the four-form field strength  is now:
\beq
F\,=\,\mu\,dx^+\wedge dy^1\wedge dy^2\wedge dy^3\,+\,
dx^+\wedge dx^-\wedge dy^1\wedge dH^{-1}\,\,.
\label{nuno}
\eeq
Notice that the Bianchi identity $dF=0$ is automatically satisfied. Moreover,
the wave term in $F$ gives rise to the following component of the Hodge dual
field strength:
\beq
{}^*F_{x^+z^1\cdots z^6}\,=\,\mu\,H\,.
\label{ndos}
\eeq
Since, $F\wedge F=0$, the field equation (\ref{snueve}) reduces to
 $d{}^*F=0$. By inspecting the component (\ref{ndos}) of ${}^*F$
one arrives at the conclusion that $H$ must depend
only on the $z$ coordinates, \ie:
\beq
H=H(\,\vec z\,)\,\,.
\label{ntres}
\eeq
Thus our M2-brane is smeared in the $(y^2,y^3)$ directions. The full expression
of ${}^*F$ is:
\beq
{}^*F\,=\,\mu\,H\,dx^+\wedge dz^1\wedge\cdots \wedge dz^6\,-\,
dy^2\wedge dy^3\wedge {}^{\widetilde*}dH\,\,,
\label{ncuatro}
\eeq
where ${}^{\widetilde*}$ denotes now the Hodge dual with respect to the
coordinates $z^1\cdots z^6$ with the Euclidean metric. The equation  $d{}^*F=0$
for the second term in ${}^*F$ implies that $H$ must be a harmonic
function of
$z^1\cdots z^6$. Let us write it as:
\beq
H\,=\,1\,+\,{Q\over |\,\vec z\,\,|^4}\,\,,
\label{ncinco}
\eeq
where $Q$ is a constant related to the charge of the M2-brane. To determine
completely the metric, let us write the profile equation (\ref{noventa}) for
this $m=1$ case:
\beq
\partial_{y^1}^2\,W\,+\,H^{-1}\,
\big[\,\partial^2_{\tilde y^{\,\alpha}}\,
+\,\partial^2_{\tilde z^{\,\alpha}}\,\big]\,W\,=\,-\mu^2\,\,.
\label{nseis}
\eeq
In order to solve this equation, let us represent $W$ as:
\beq
W\,=\,-\Big(\,{\mu\over 3}\,\Big)^2\,\vec y^{\,2}\,-\,
\Big(\,{\mu\over 6}\,\Big)^2\,\vec z^{\,2}\,+\,f(\vec z)\,\,,
\label{nsiete}
\eeq
where $f(\vec z)$ is a function to be determined. Notice that the ansatz
(\ref{nsiete}) ensures that $f=0$ for $Q=0$. By plugging the expression
(\ref{ncinco}) for $H$ and our ansatz (\ref{nsiete}) for $W$ in eq.
(\ref{nseis}), one gets that $f(\vec z)$ satisfies the equation:
\beq
\partial_{\vec z}^2\,f\,=\,-{7\over 9}\,\mu^2\,{Q\over |z|^4}
\label{nocho}
\eeq
This type of equation has been solved in general in appendix B.
Particularizing to the case of eq.
(\ref{nocho}), we conclude that, up to a harmonic function,  $f$ is:
\beq
f\,=\,{7\over 36}\,\mu^2\,Q{1\over |z|^2}\,\,.
\label{nnueve}
\eeq
Then, the full profile for the $(+,-,1,0)$ configuration is:
\beq
W\,=\,-\Big(\,{\mu\over 3}\,\Big)^2\,\vec y^{\,2}\,-\,
\Big(\,{\mu\over 6}\,\Big)^2\,\vec z^{\,2}\,+\,
{7\over 36}\,\mu^2\,Q\,\,{1\over \vec z^{\,2}}\,\,.
\label{cien}
\eeq
Notice that, at large distances from the brane, the back-reaction term $f$ is
subleading with respect to the one corresponding to the pure pp-wave. This, as
we will verify case by case, is a general fact for the solutions we will
obtain.

We have already argued in general that M-branes in this pp-wave can only have
standard spinors. For this $(+,-,1,0)$ embedding $\theta=\mu I$, which commutes
with $\Gamma_{\kappa}$ and is independent of the external coordinates.
Thus, the
two conditions of eq. (\ref{odos}) are satisfied and  we have eight
standard spinors of the form displayed in eq. (\ref{otres}).

\subsection{M5 branes}
\medskip
For a $(+,-,m,n)$ M5-brane configuration the wave contribution to $F$ in
eq. (\ref{osiete}) gives
rise to the following component of ${}^*F$:
\beq
{}^*F_{x^+z^1\cdots z^6}\,=\,\mu\,H^{m-1}\,\,.
\label{ctuno}
\eeq
Thus,  for $m\not=1$, the equation $d{}^*F=0$ implies that $H$ must be
independent of the $y$ coordinates and, therefore:
\beq
H\,=\,H(\,\tilde z^{\alpha}\,)\,\,.
\label{ctdos}
\eeq
We have seen in section 2.1.2 that $\Gamma_{\kappa}$ commutes with $I$ only for
$m=0,2$. Then, from now on, we will only consider these two cases, for which
eq. (\ref{ctdos}) must hold. Notice that this means that our branes must be
smeared along the external $y$ directions. The full ansatz for $F$ will be:
\beq
F\,=\,\mu\,dx^+\wedge dy^1\wedge dy^2\wedge dy^3\,+\,
{}^{\widetilde*}dH\wedge dy^{m+1}\wedge\cdots \wedge dy^3\,\,,
\label{cttres}
\eeq
where again ${}^{\widetilde*}$ denotes the Hodge dual with respect to the
$\tilde z^{\alpha}$ coordinates with the Euclidean metric. Notice that
$F\wedge F=0$ for $m=0,2$. The Bianchi identity
is now non-trivial and imposes that $H$ is a harmonic function of the
$\tilde z$
coordinates:
\beq
\partial_{\tilde z^{\alpha}}^2\,H\,=\,0\,\,.
\label{ctcuatro}
\eeq
Moreover, the full expression for the Hodge dual is now:
\bear
{}^*F&=&\mu\,H^{m-1}\,dx^+\wedge dz^1\wedge\cdots \wedge dz^6\,+\,\rc\rc
&&+\,dx^+\wedge dx^-\wedge dy^1\wedge\cdots \wedge dy^m
\wedge dz^1\wedge\cdots \wedge dz^n\wedge dH^{-1}\,\,,
\label{ctcinco}
\eear
and the equation of motion is satisfied  as a consequence of eq.
(\ref{ctdos}).

For these $(+,-,m,n)$ M5-brane configurations the matrix $\theta$ is given by:
\beq
\theta\,=\,\mu\,H^{{m-2\over 2}}\,
\Gamma_{\,\widehat y^{\,1}\,\widehat y^{\,2}\,\widehat y^{\,3}}\,\,,
\label{ctseis}
\eeq
and, therefore,  the profile equation becomes:
\beq
\partial_{a}^2\,W\,+\,H^{-1}\,\partial_{\alpha}^2\,W\,=\,-
\mu^2\,H^{m-2}\,\,.
\label{ctsiete}
\eeq
In what follows we will analyze separately the $m=0$ and $m=2$ cases. Notice,
however, that only for $m=2$ the matrix $\theta$ in eq. (\ref{ctseis}) is
independent of the external coordinates and, thus, only for this case the
corresponding supergravity solution is supersymmetric.

\subsubsection{$(+,-,2,2)$}
\medskip
According to eq. (\ref{ctdos}), the harmonic function $H$ will only depend on
the four external $z$ coordinates
$\tilde z^{\alpha}\,=\,(\,z^3,\cdots,z^6\,)$. Thus, we can write:
\beq
H\,=\,1\,+\,{Q\over |\,\tilde z\,|^2}\,\,.
\label{ctocho}
\eeq
Moreover, it follows from eq. (\ref{ctsiete}) that the profile equation in this
case becomes:
\beq
\big[\,\partial^2_{y^a}\,+\,\partial^2_{z^a}\,\big]\,W\,+\,
H^{-1}\,
\big[\,\partial^2_{\tilde y^{\,\alpha}}\,
+\,\partial^2_{\tilde z^{\,\alpha}}\,\big]\,W\,=\,-\mu^2\,\,.
\label{ctnueve}
\eeq
Let us try to find a solution to equation (\ref{ctnueve}) of the form:
\beq
W\,=\,-\Big(\,{\mu\over 3}\,\Big)^2\,\vec y^{\,2}\,-\,
\Big(\,{\mu\over 6}\,\Big)^2\,\vec z^{\,2}\,+\,f(\tilde z)\,\,,
\label{ctdiez}
\eeq
where we have assumed that the unknown function $f$ depends on the
same variables as the harmonic function $H$ in eq. (\ref{ctocho}). After
substituting the ansatz (\ref{ctdiez}) in (\ref{ctnueve}),
one arrives at the following differential equation for  $f(\tilde z)$:
\beq
\partial^2_{\tilde z^{\,\alpha}}\,f\,=\,-{4\mu^2\over 9}\,
{Q\over |\,\tilde z\,|^2}\,\,,
\label{ctonce}
\eeq
whose  solution  can be obtained from the results of appendix B, namely:
\beq
f\,=\,-{1\over 9}\,\mu^2 Q\,\log (\,\tilde z^{\,2}\,)\,\,.
\label{ctdoce}
\eeq
Therefore, the brane contribution to the profile is, in this case, a
logarithmic function. Actually, the full profile is given by:
\beq
W\,=\,-\Big(\,{\mu\over 3}\,\Big)^2\,\vec y^{\,2}\,-\,
\Big(\,{\mu\over 6}\,\Big)^2\,\vec z^{\,2}\,-\,
{1\over 9}\,\mu^2 Q\,\log (\,\tilde z^{\,2}\,)\,\,,
\label{cttrece}
\eeq
and, as this solution satisfies eq. (\ref{odos}), it has eight standard spinors
of the form (\ref{otres}).

\subsubsection{$(+,-,0,4)$}
\medskip
The external $z$ coordinates in this case can be taken as
$\tilde z^{\alpha}\,=\,(\,z^5,z^6\,)$ and the solution of eq.
(\ref{ctcuatro}) gives a harmonic function which is logarithmic in the
$\tilde z^{\alpha}$ coordinates:
\beq
H\,=\,1\,+\,Q\,\log(\,\tilde z^2\,)\,\,.
\label{ctcatorce}
\eeq
Moreover, the profile equation (\ref{ctsiete}) for this case is:
\beq
\partial^2_{z^a}\,W\,+\,
H^{-1}\,
\big[\,\partial^2_{\tilde y^{\,\alpha}}\,
+\,\partial^2_{\tilde z^{\,\alpha}}\,\big]\,W\,=\,-H^{-2}\,\mu^2\,\,.
\label{ctquince}
\eeq
Due to the presence of the logarithm in the expression of $H$, the solution of
(\ref{ctquince}) is not obtainable in terms of elementary functions and we will
not try to find it. Notice that now $\theta=\mu H^{-1} I$ (see eq.
(\ref{ctseis})) and, thus, the second equation in (\ref{odos}) cannot be
satisfied. Therefore, this case is not supersymmetric unless we smear
completely
the brane by taking $H$ constant which gives rise, after some coordinate
redefinition, to the original pp-wave.

\setcounter{equation}{0}
\section{M-branes in the 24-SUSY pp-wave}
\medskip
Let us split the transverse coordinates $x^i$ as in section 2.2, namely
$x^i=(\vec y, \vec z, x^{9})$, where $\vec y$ and $\vec z$ are vectors with
four components. Sometimes it will be useful to differentiate between
coordinates parallel and orthogonal to the brane worldvolume. We will use the
same conventions as in previous sections, namely, the coordinates along the
brane worldvolume will be labeled by a latin index, whereas those transverse
to the brane will have greek indices and a tilde. The four-form gauge field
strength for the intersection of an M-brane and the pp-wave with 24
supersymmetries will be of the form:
\beq
F\,=\,\mu\,\big[\,dx^{+}\wedge dy^{1}\wedge  dy^{2}
\wedge dx^{9}\,+\,
dx^{+}\wedge dy^{3}\wedge  dy^{4}\wedge dx^{9}
\,\big]\,+\,{\rm brane\,\,\, term}\,\,.
\label{ctdseis}
\eeq

It is rather easy to reach the conclusion that the only configurations
which can be supersymmetric are those for which $\Gamma_{\kappa}$ commutes or
anticommutes with $\theta$. Indeed, $\theta$ is now the sum of two terms
$\theta=\theta_1+\theta_2$, where $\theta_1$ and $\theta_2$ are proportional to
a single product of transverse $\Gamma$-matrices. Thus $\theta_1$ and
$\theta_2$ cannot have zero modes. Let us assume, say, that
$[\Gamma_{\kappa},\theta_1]=0$ and $\{\Gamma_{\kappa},\theta_2\}=0$. We will
now prove that there are not Killing spinors in this case. First of all, the
first condition in (\ref{odos}) implies that $\epsilon^{st}$ must be a zero
mode of $\theta_2$, which is impossible. This excludes the possibility of
having standard Killing spinors. On the other hand, the condition
$[\Gamma_{\kappa},\Omega_a^{w}]\epsilon=0$ of eq. (\ref{ocuatro}) implies
$\theta_2\Gamma_{\widehat -}\,\chi=0$, which  cannot be satisfied by
supernumerary Killing spinors (for which  $\Gamma_{\widehat -}\,\chi\not=0$).
This reduces our analysis to the configurations listed in (\ref{sesenta}) and
(\ref{stres}) for the M2 and M5 branes respectively. In each case we have
to study the fulfillment of eqs. (\ref{odos}) and (\ref{oseis}) for
standard and supernumerary spinors respectively. At this point it is
interesting to recall that the $\Omega_{y}$ matrices do not  have
supernumerary zero modes which, in most of the  embeddings,  excludes the
possibility of having supernumerary Killing spinors. As in the maximally
SUSY case, we will proceed through a case by
case study.

\subsection{M2-branes }
\medskip
For a M2-brane embedding of the type $(+,-, (m_1,m_2), n, p)$, the $\theta$
matrix is:
\beq
\theta\,=\,\mu\,\big[\,H^{{m_1+p-1\over 2}}\,\,
\Gamma_{\,\widehat y^{\,1}\,\widehat y^{\,2}\,\widehat x^{\,9}}\,+\,
H^{{m_2+p-1\over 2}}\,\,
\Gamma_{\,\widehat y^{\,3}\,\widehat y^{\,4}\,\widehat x^{\,9}}\,
\big]\,\,,
\label{ctdsiete}
\eeq
and, therefore,  the profile equation becomes:
\beq
\partial_{a}^2\,W\,+\,H^{-1}\,\partial_{\alpha}^2\,W\,=\,-
\mu^2\,\big[\,H^{m_1+p-1}\,+\,H^{m_2+p-1}\,\big]\,\,.
\label{ctdocho}
\eeq

\subsubsection{ $(+,-,0,1,0)$  }
\medskip
This case corresponds to taking $m_1=m_2=p=0$. By computing the Hodge dual of
the wave term in (\ref{ctdseis}) one can prove that ${}^*\,F$
has the following components:
\beq
(\,{}^*\,F\,)_{x^+z^1\cdots z^4 y^1y^2}\,=\,
(\,{}^*\,F\,)_{x^+z^1\cdots z^4 y^3y^4}\,=\,\mu\,\,.
\label{ctdnueve}
\eeq
Notice that no power of $H$ appears on the right-hand side of eq.
(\ref{ctdnueve}). As a consequence the components of (\ref{ctdnueve}) do not
contribute to $d{}^*\,F$ and there will be no need of smearing the M2-brane.
Without loss of generality we shall extend the M2-brane along the $z^1$
direction. Accordingly, the brane term of $F$ is given by:
\beq
F_{M2}\,=\,dx^+\wedge\ dx^-\wedge dz^1\wedge dH^{-1}\,\,.
\label{ctveinte}
\eeq
Then, it follows from eqs. (\ref{ctdseis}) and (\ref{ctveinte}) that
$dF=0$ automatically and, by computing ${}^*\,F_{M2}$, one can check that
$d{}^*\,F=0$ if $H$ is a harmonic function
of all the eight external  coordinates $y\,=\,(\,y^1\,,\cdots\,,y^4\,)$,
$\tilde z\,=\,(z^2,z^3,z^4)$ and $x^9$:
\beq
H\,=\,1\,+\,{Q\over [\,y^2+\tilde z^2+\,(x^9)^2\,]^3}\,\,.
\label{ctvuno}
\eeq
Moreover, the profile equation in this case is:
\beq
\partial_{a}^2\,W\,+\,H^{-1}\,\partial_{\alpha}^2\,W\,=\,-2\mu^2\,H^{-1}\,\,,
\label{ctvdos}
\eeq
and is solved by taking:
\beq
W\,=\,-{\mu^2\over 4}\,y^2\,\,.
\label{ctvtres}
\eeq
Notice that there is no correction with respect to the pure pp-wave term.

Let us now study the supersymmetries of this embedding. Notice, first of all,
that $\theta=\mu\, H^{-{1\over 2}}\,J$ for this case (see eq. (\ref{ctdsiete}))
and that $\{\Gamma_{\kappa}, \theta\}=0$. Then, in order to satisfy eq.
(\ref{odos}) we have to require that $\epsilon^{st}$ be a zero mode of
$\theta$. Thus, the standard Killing spinors are of the form displayed in eq.
(\ref{otres}) with the extra condition $J\chi^{st}=0$, which gives four of
them. Moreover, from the requirements of (\ref{oseis}) we conclude that the
supernumerary Killing spinors must be annihilated by $\Omega_{z^1}$, which is
only possible if they are also annihilated by $J\Gamma_{\widehat -}$. This, in
turn, implies that
$\Omega_{z^2}\epsilon=\Omega_{z^3}\epsilon=\Omega_{z^4}\epsilon=0$. With this
information, and the explicit form of the matrices $\theta$ and
$\Omega_{y^i}$, it is easy to integrate the system (\ref{ochenta}) for
supernumerary Killing spinors. The result is just the one written in eqs.
(\ref{cicuatro}) and (\ref{cicinco}), with the extra condition
$\Gamma_{\kappa}\chi^{sn}=\chi^{sn}$, which gives four supernumerary spinors.
Thus, summing up, this system is
$8(4+4)$ supersymmetric.

\medskip
\subsubsection{ $(+,-,0,0,1)$  }
\medskip

The M2-brane in this case is extended along the $x^9$ direction and, therefore,
the brane term in the four-form is:
\beq
F_{M2}\,=\,dx^+\wedge\ dx^-\wedge dx^9\wedge dH^{-1}\,\,.
\label{ctvcuatro}
\eeq
As now the wave contribution to $F$ gives rise to the following components of
${}^*\,F$:
\beq
(\,{}^*\,F\,)_{x^+z^1\cdots z^4 y^1y^2}\,=\,
(\,{}^*\,F\,)_{x^+z^1\cdots z^4 y^3y^4}\,=\,\mu\,H\,\,,
\label{ctvcinco}
\eeq
then, $d{}^*\,F=0$ if the brane is smeared in the $y$  directions and $H$ is a
harmonic function of the $z$ coordinates. Thus:
\beq
H\,=\,1\,+\,{Q\over z^2}\,\,.
\label{ctvseis}
\eeq
Moreover, the profile equation:
\beq
\partial_{a}^2\,W\,+\,H^{-1}\,\partial_{\alpha}^2\,W\,=\,-2\mu^2\,\,,
\label{ctvsiete}
\eeq
is solved by the following function:
\beq
W\,=\,-{\mu^2\over 4}\,y^2\,-\,{\mu^2 Q\over 2}\,\log (z^2)\,\,.
\label{ctvocho}
\eeq
For this embedding $\theta=\mu J$ (see eq. (\ref{ctdsiete})) and
$\Gamma_{\kappa}$ commutes with $\theta$. Therefore,  eq. (\ref{odos}) is
automatically satisfied. On the other hand, eq. (\ref{oseis}) implies, in
particular, that $\Omega_{y^i}\epsilon=0$, which is not possible  for
spinors with $\Gamma_{\widehat -}\,\epsilon\not=0$. Thus, there are no
supernumerary Killing spinors and this configuration is $8(8+0)$
supersymmetric.

\subsection{M5-branes }
\medskip
Let us consider a M5-brane embedding of the type $(+,-, (m_1,m_2), n, p)$.
By computing with the metric of this configuration the contribution to
${}^*\,F$
of the wave term in  (\ref{ctdseis}), one gets:
\bear
(\,{}^*\,F\,)_{x^+z^1\cdots z^4 y^1y^2}&=&\mu\,H^{m_2+p-1}\,\,,\rc\rc
(\,{}^*\,F\,)_{x^+z^1\cdots z^4 y^3y^4}&=&\mu\,H^{m_1+p-1}\,\,.
\label{ctvnueve}
\eear
The study of the powers of $H$ on the right-hand side of eq. (\ref{ctvnueve})
for the different cases will allow us to determine the precise form of $H$.
Moreover, the matrix $\theta$ is now given by:
\beq
\theta\,=\,\mu\,\big[\,H^{{m_1+p-2\over 2}}\,\,
\Gamma_{\,\widehat y^{\,1}\,\widehat y^{\,2}\,\widehat x^{\,9}}\,+\,
H^{{m_2+p-2\over 2}}\,\,
\Gamma_{\,\widehat y^{\,3}\,\widehat y^{\,4}\,\widehat x^{\,9}}\,
\big]\,\,,
\label{cttreinta}
\eeq
and, therefore, the profile equation takes the form:
\beq
\partial_{a}^2\,W\,+\,H^{-1}\,\partial_{\alpha}^2\,W\,=\,-\mu^2\,
\big(\,H^{m_1+p-2}\,+\,H^{m_2+p-2}\,\big)\,\,.
\label{cttuno}
\eeq

\medskip
\subsubsection{ $(+,-,(2,2),0,0)$  }
\medskip
By inspecting (\ref{ctvnueve}) one readily realizes that
in this case $H$ should be independent of $x^9$ and  harmonic on  the
four $z$ coordinates, \ie:
\beq
H=1+{Q\over |z|^2}\,\,,
\label{cttdos}
\eeq
while the profile  equation (\ref{cttuno}) for $m_1=m_2=2$ and $p=0$ is solved
by:
\beq
W\,=\,-{\mu^2\over 4}\,y^2\,\,.
\label{ctttres}
\eeq
Moreover, since now $\theta=\mu J$ and $[\Gamma_{\kappa},\theta]=0$, the
conditions (\ref{odos}) are trivially satisfied and we have eight standard
Killing spinors of the form (\ref{otres}). On the other hand, it follows from
(\ref{oseis}) that the Killing spinors must be independent of the external
coordinates $z^i$ and $x^9$. Actually, it is not difficult to demonstrate that
this configuration has four supernumerary spinors of the type (\ref{cicuatro}),
where, in addition to (\ref{cicinco}), $\chi^{sn}$ satisfies the condition
$\Gamma_{\kappa}\chi^{sn}=\chi^{sn}$. All together this configuration has
$12(8+4)$ supersymmetries. This system has also been  studied in ref.
\cite{Singh}.

\medskip
\subsubsection{ $(+,-,(2,0),2,0)$  }
\medskip
The external coordinates in this case are
$(\tilde y\,,\,\tilde z\,,\,x^9)\,=\,(\,y^3,y^4,z^3,z^4,x^9\,)$. Since
$m_2+p-1\,=\,-1$, $H$ should not depend on $\tilde y\,=\,(\,y^3,y^4\,)$ and on
$x^9$. Therefore $H$ only depends on $\tilde z\,=\,(\,z^3,z^4\,)$ in the form:
\beq
H\,=\,1\,+\,Q\log (\tilde z^2)\,\,.
\label{cttcuatro}
\eeq
The profile equation cannot be solved in terms of elementary functions.
Actually, since now
$\theta\,=\,\mu\,\big[\,
\Gamma_{\,\widehat y^{\,1}\,\widehat y^{\,2}\,\widehat x^{\,9}}\,+\,
H^{-1}\,\,
\Gamma_{\,\widehat y^{\,3}\,\widehat y^{\,4}\,\widehat x^{\,9}}\,
\big]$, the second equation in (\ref{odos}) cannot be satisfied and there are
no standard spinors. It might be equally verified that supernumerary spinors
cannot exist and, thus, this supergravity solution is not supersymmetric
(unless we put $Q=0$, which is the original pp-wave).

\medskip
\subsubsection{ $(+,-,(0,0),4,0)$  }
\medskip
The external coordinates are now the four $y$'s and $x^9$. Since
$m_1+p-1=m_2+p-1=-1$, the harmonic function should not depend on
$y$ and $x^9$, \ie\ it must be a constant. Thus, the brane contribution  to $F$
vanishes and we obtain the pure pp-wave solution with a redefinition of $\mu$.

\medskip
\subsubsection{ $(+,-,(1,1),1,1)$  }
\medskip

In this case $\tilde y=(y^2,y^4)$, $\tilde z=(z^2,z^3,z^4)$ and, since
$m_1+p-1=m_2+p-1=1$, $H$ is independent of $\tilde y$ and equal to:
\beq
H\,=\,1\,+\,{Q\over |\tilde z|}\,\,,
\label{cttcinco}
\eeq
Moreover, the profile can be taken as:
\beq
W\,=\,-{\mu^2\over 4}\,y^2\,-\,{Q\mu^2\over 2}\,\,
|\tilde z|\,\,.
\label{cttseis}
\eeq
Concerning supersymmetry, as $\theta=\mu J$ and $[\Gamma_{\kappa}, \theta]=0$,
the conditions (\ref{odos})  are trivially satisfied and we have eight standard
spinors. Moreover, eq. (\ref{oseis}) requires the existence of zero modes of
the $\Omega_{\tilde y}$ matrices, which cannot occur for supernumerary spinors.
Therefore this configuration is $8(8+0)$ supersymmetric.

\medskip
\subsubsection{ $(+,-,(1,1),2,0)$  }
\medskip

Now $\tilde y=(y^2,y^4)$, $\tilde z=(z^3,z^4)$ and, as
$m_1+p-1=m_2+p-1=0$, $H$ must depend on $\tilde y$, $\tilde z$ and $x^9$ as
follows:
\beq
H\,=\,1\,+\,
{Q\over [\,\tilde y^{\,2}+ \tilde z^{\,2}+(x^9)^2\,]^{{3\over 2}}}\,\,,
\label{cttsiete}
\eeq
and the profile $W$ is:
\beq
W\,=\,-{\mu^2\over 4}\,y^2\,-\,{Q\mu^2\over 2}\,\,
{1\over [\,\tilde y^{\,2}+ \tilde z^{\,2}+(x^9)^2\,]^{{1\over 2}}}\,\,.
\label{cttocho}
\eeq
For this embedding $\theta=\mu H^{-{1\over  2}} J$ and, since
$\{\Gamma_{\kappa}, \theta\}=0$, one concludes  from (\ref{odos}) that there
are four standard Killing spinors as those written in (\ref{otres}) with
$J\chi^{st}=0$. Moreover, it is straightforward to prove that there are no
supernumerary Killing spinors and, thus, this configuration is $4(4+0)$
supersymmetric.

\medskip
\subsubsection{ $(+,-,(2,0),1,1)$  }
\medskip
Now $m_2+p-1=0$ and, thus, by requiring that $d{}^*F=0$, one concludes that
$H$ should  depend on $\tilde y=(y^3,y^4)$ and
$\tilde z=(z^2,z^3,z^4)$. However, one can check that
$F_{wave}\wedge F_{M5}$ is not zero and, therefore, the condition  $d{}^*F=0$
does not guarantee that the equation of motion of $F$  are satisfied (this does
not happen in the cases studied so far). Then, in this case we are not
able even to solve the supergravity equations of motion for non-trivial $H$.

\medskip
\subsubsection{ $(+,-,(0,0),3,1)$  }
\medskip

In this case, $\tilde y=y$, $\tilde z=z^4$ and, since
$m_1+p-1=m_2+p-1=0$, $H$ will depend on $y$ and $z^4$:
\beq
H\,=\,1\,+\,{Q\over [\, y^{\,2}+  (z^4)^{2}\,]^{{3\over 2}}}\,\,,
\label{cttnueve}
\eeq
and from the profile equation it follows that one can take:
\beq
W\,=\,-{\mu^2\over 4}\,y^2\,\,.
\label{ctcuarenta}
\eeq
Now $\theta=\mu H^{{1\over 2}} J$ and $\Gamma_{\kappa}$ anticommutes with
$\theta$. Thus, we will satisfy (\ref{odos}) if we require, in addition to the
requirements of (\ref{otres}), that
$J\chi^{st}=0$, which gives four standard spinors. The conditions
(\ref{oseis}) can be satisfied by supernumerary spinors and, actually, one can
easily integrate the corresponding system of differential equations
(\ref{ochenta}). The result are just the spinors displayed in eq.
(\ref{cicuatro}), with the additional condition
$\Gamma_{\kappa}\chi^{sn}=\chi^{sn}$, which restricts the number of
supernumerary spinors to be one half of those of the pure pp-wave, \ie\
four. In
conclusion this system is $8(4+4)$ supersymmetric.

\medskip
\section{Summary and Conclusions}
\medskip

In this paper we have studied supersymmetric intersections of branes and
pp-waves in M-theory. We have first looked at this problem by considering brane
probes extended along fixed hyperplanes in the pp-wave background geometry, and
by using kappa symmetry to determine the number of supersymmetries preserved by
the different embeddings. This analysis leads to a series of algebraic
conditions to be satisfied by the Killing spinors of the background. We have
performed a case by case analysis and we have  determined how many standard and
supernumerary spinors satisfy these algebraic conditions for M2 and M5 branes
in the pp-wave backgrounds with 32, 24 and 20 supersymmetries.

Furthermore, we
have obtained supergravity solutions representing the wave-brane intersection,
and we have determined the number of supersymmetries they preserve. The metric
of these solutions is a warped version of that of the pp-wave, with a profile
which is generically different from that of the pure pp-wave case. Moreover,
the fulfillment of the equations of motion of the four-form requires in many
cases that the M-branes be delocalised along some directions of the four-form
pp-wave flux.

In general, the requirements imposed by supersymmetry in
the supergravity analysis are more restrictive that those found in the brane
probe approach. Due to the delocalisation of the solution, one expects to make
contact with the case of the brane probe outside the origin. This happens in
most of the cases, except in some ones in which the supersymmetry is completely
lost in the supergravity solution, due to the presence of the harmonic function
in some terms of the Killing spinor equation. It is also interesting to point
out that there are very few cases in which supernumerary spinors survive
at the level of the supergravity analysis. All these embeddings share the
distinguishing feature that they present no deformation of the
wave profile. This fact points towards the possibility of  obtaining these
configurations as  Penrose limits of non-standard intersections
\cite{Singh}.

Our analysis has been systematic but by no means completely exhaustive. In the
case of the M5 brane probe, for example, one could try to switch on
worldvolume gauge fields that change the kappa symmetry matrix and could make
some embeddings supersymmetric. This is actually what happens in the type IIB
analysis of ref. \cite{Sken} and in the $(+,-,2,2)$ configurations in the
maximally SUSY eleven dimensional pp-wave \cite{Kim}. On the other hand, it was
claimed in ref. \cite{Skendos} that some broken  spacetime supersymmetries in
the type IIB theory are restored by using worldsheet symmetries. It would be
interesting to find an eleven dimensional analogue of this phenomenon.
Another possibility is to consider spherical
branes. From the matrix theory approach it is known that there are
supersymmetric configurations of this kind, \ie\ fuzzy spheres, which can be
traced back to the giant gravitons of the $AdS_{4,7}\times S^{7,4}$ space. More
generally one could also try to find worldvolume solitons (bions) on the
pp-wave background.

On the supergravity side, it would be worth to reconsider those cases for which
the brane probe approach predicts the existence of a supersymmetric embedding
and, however, we have failed in finding a (supersymmetric) solution of the
supergravity equations. In these cases we would have to explore the possibility
of modifying our general ansatz (notice, for example,  that fully localized
intersections of pp-waves and D-branes were constructed in ref. \cite{Alday}).
Finally it would be also interesting to reduce our solutions to ten dimensions
and apply them the different string theory dualities. Notice that most of our
solutions have isometries which allow this dimensional reduction. In this way
one expects to find solutions representing branes in G\"odel universes
\cite{Godel}.

We expect to report on some of these issues elsewhere.

\medskip
\section*{Acknowledgments}
\medskip
We are grateful to Joaquim Gomis for collaboration in the initial stages
of this project.  We would like also to thank Angel Paredes for a critical
reading of the manuscript.  This work has been supported in part by MCyT and
FEDER under grant  BFM2002-03881 and by  the EC Commission under the FP5
grant HPRN-CT-2002-00325.

\vskip 1cm
\renewcommand{\theequation}{\rm{A}.\arabic{equation}}
\setcounter{equation}{0}
\medskip
\appendix
\section{Ricci tensor for wave-brane intersections}
\medskip
Let us consider a metric in $D$ dimensions of the form:
\beq
ds^2_D\,=\,h_1\,(\,2dx^+dx^-\,+\,W\,(dx^+)^2\,+(dx^a)^2\,)\,+\,
h_2\,(\,d\tilde x^{\,\alpha}\,)^2\,\,,
\label{apauno}
\eeq
where $x^a\,=\,(x^1,\cdots,x^{p-1}\,)$ and $h_1$ and $h_2$ depend on the
coordinates $\tilde x^{\,\alpha}$, whereas $W$ is a function of both $x^a$ and
$\tilde x^{\,\alpha}$. We consider the following basis of one-forms:
\bear
e^{\widehat +}&=&h_1^{{1\over 2}}\,dx^+\,,
\,\,\,\,\,\,\,\,\,\,\,\,\,\,\,\,\,
e^{\widehat -}\,=\,h_1^{{1\over 2}}(\,dx^-\,+\,{W\over 2}\,dx^+\,)\,\,,\rc\rc
e^{\widehat a}&=&h_1^{{1\over 2}}\,dx^a\,,
\,\,\,\,\,\,\,\,\,\,\,\,\,\,\,\,\,
e^{\widehat \alpha}=h_2^{{1\over 2}}\,d\tilde x^{\,\alpha}\,\,.
\label{apados}
\eear
In this basis $ds^2_D\,=\,2e^{\widehat +}\,e^{\widehat -}\,+\,
e^{\widehat a}\,e^{\widehat a}\,+\,e^{\widehat \alpha}\,e^{\widehat
\alpha}$. The spin
connection is:
\bear
\omega^{\widehat +\,\widehat \alpha}&=&{1\over
2}\,\big(\,h_1h_2\,\big)^{-{1\over
2}}
\,\partial_{\alpha}h_1\,dx^+\,\,,\rc\rc
\omega^{\widehat -\,\widehat \alpha}
&=&{1\over 2}\,\big(\,h_1h_2\,\big)^{-{1\over 2}}\,
\partial_{\alpha}h_1\,(\,dx^-\,+\,{W\over 2}\,dx^+\,)\,+\,
{1\over 2}\,\Bigg(\,{h_1\over h_2}\,\Bigg)^{{1\over
2}}\,\partial_{\alpha}\,W\,dx^+\,\,,\rc\rc
\omega^{\widehat -\,\widehat a}&=&{1\over 2}\,\partial_a W\,dx^+\,\,,\rc
\omega^{\widehat a\,\widehat \alpha}
&=&{1\over 2}\,\big(\,h_1h_2\,\big)^{-{1\over 2}}
\,\partial_{\alpha}h_1\,dx^a\,\,,\rc\rc
\omega^{\widehat \alpha\,\widehat \beta}&=&{1\over 2}\,h_2^{-1}\,\,
(\partial_{\beta}h_2\,d\tilde x^{\,\alpha}\,-\,
\partial_{\alpha}h_2\,d\tilde x^{\,\beta}\,\,)\,\,.
\label{apatres}
\eear
The light-cone components of the Ricci tensor are:
\bear
R_{\widehat-\,\widehat-}&=&0\,,\rc\rc
R_{\widehat+\,\widehat+}&=&-{1\over 2}\,h_1^{-1}\,\partial_a^2\,W\,-\,
{1\over 2}\,h_2^{-1}\,\partial_{\alpha}^2\,W\,+\,\rc\rc
&&+\,h_2^{-1}\,\Big[\,-{p+1\over 4}\,\partial_{\alpha}\,\log h_1\,+\,
{p-D+3\over 4}\,\partial_{\alpha}\,\log h_2\,\Big]\,
\partial_{\alpha}\,W\,\,,\rc\rc
R_{\widehat+\,\widehat-}&=&-{1\over 2}\,\big(\,h_1h_2\,\big)^{-1}\,
\partial_{\alpha}^2\,h_1\,+\,\rc\rc
&&+\,h_2^{-1}\,\Big[\,{1-p\over 4}\,\partial_{\alpha}\,\log h_1\,+\,
{p-D+3\over 4}\,\partial_{\alpha}\,\log h_2\,\Big]\,
\partial_{\alpha}\log h_1\,\,.\rc
\label{apacuatro}
\eear
The  components of the Ricci tensor along the worldvolume of the brane
are:
\bear
R_{\widehat a\,\,\widehat b}&=&\delta_{ab}\,
\Big[\,-{1\over 2}\,\big(\,h_1h_2\,\big)^{-1}\,
\partial_{\alpha}^2\,h_1\,+\,\rc\rc
&&+\,h_2^{-1}\,\Big(\,{1-p\over 4}\,\partial_{\alpha}\log h_1\,+\,
{p-D+3\over 4}\,\partial_{\alpha}\,\log h_2\,\Big)\,
\partial_{\alpha}\log h_1\,\Big]\,\,.\rc
\label{apacinco}
\eear
In order to write the components of the Ricci tensor orthogonal to the brane,
let us define:
\beq
\varphi\,=\,{1\over 2}\,(p+1)\,\log h_1\,+\,
{1\over 2}\,(D-p-3)\,\log h_2\,\,.
\label{apaseis}
\eeq
Then:
\bear
R_{\widehat \alpha\,\,\widehat\beta}&=&h_2^{-1}\,\Big[\,
-\partial_{\alpha}\partial_{\beta}\,\varphi\,+\,
{1\over 2}\,\partial_{\alpha}\log h_2\,\partial_{\beta}\,\varphi\,+\,
{1\over 2}\,\partial_{\beta}\log h_2\,\partial_{\alpha}\,\varphi\,-\rc\rc
&&-\,{p+1\over 4}\,\partial_{\alpha}\log h_1\,\partial_{\beta}\log h_1\,-\,
{D-p-3\over 4}\,\partial_{\alpha}\log h_2\,\partial_{\beta}\log h_2\,-\,\rc\rc
&&-\,{\delta_{\alpha\beta}\over 2}\,\partial^2_{\gamma}\,\log h_2\,-\,
{\delta_{\alpha\beta}\over 2}\,\partial_{\gamma}\,\log h_2\,
\partial_{\gamma}\,\varphi\,\Big]\,\,.
\label{apasiete}
\eear
For a metric of M2 type we put $D=11$, $p=2$, $h_1=H^{-2/3}$ and
$h_2=H^{1/3}$. We get:
\bear
R_{\widehat +\,\widehat+}&=&-{1\over 2}\,H^{{2\over 3}}\,\partial^2_a\,W
\,-\,{1\over 2}\,H^{-{1\over 3}}\,\partial^2_{\alpha}\,W\,\,,\rc\rc
R_{\widehat +\,\widehat-}&=&-{1\over 3}\,H^{-{7\over 3}}\,
\big(\,\partial_{\alpha}\,H\,\big)^2\,+\,
{1\over 3}\,H^{-{4\over 3}}\,\partial^2_{\alpha}\,H\,\,,\rc\rc
R_{\widehat a\,\,\widehat b}&=&-{1\over 3}\,\delta_{ab}\,H^{-{7\over 3}}\,
\big(\,\partial_{\alpha}\,H\,\big)^2\,+\,
{1\over 3}\,\,\delta_{ab}\,H^{-{4\over 3}}\,\partial^2_{\alpha}\,H\,\,,\rc\rc
R_{\widehat \alpha\,\,\widehat \beta}&=&-{1\over 2}\,
H^{-{7\over 3}}\,\partial_{\alpha}H\,\partial_{\beta}H\,+\,
{\delta_{\alpha\beta}\over 6}\,\Big[\,
H^{-{7\over 3}}\,\big(\,\partial_{\gamma}\,H\,\big)^2\,-\,
H^{-{4\over 3}}\,\partial^2_{\gamma}\,H\,\Big]\,\,.
\label{apaocho}
\eear

For a metric of M5 type we put $D=11$, $p=5$, $h_1=H^{-1/3}$ and
$h_2=H^{2/3}$. We obtain:
\bear
R_{\widehat +\,\widehat+}&=&-{1\over 2}\,H^{{1\over 3}}\,\partial^2_a\,W
\,-\,{1\over 2}\,H^{-{2\over 3}}\,\partial^2_{\alpha}\,W\,\,,\rc\rc
R_{\widehat +\,\widehat-}&=&-{1\over 6}\,H^{-{8\over 3}}\,
\big(\,\partial_{\alpha}\,H\,\big)^2\,+\,
{1\over 6}\,H^{-{5\over 3}}\,\partial^2_{\alpha}\,H\,\,,\rc\rc
R_{\widehat a\,\,\widehat b}&=&-{1\over 6}\,\delta_{ab}\,H^{-{8\over 3}}\,
\big(\,\partial_{\alpha}\,H\,\big)^2\,+\,
{1\over 6}\,\,\delta_{ab}\,H^{-{5\over 3}}\,\partial^2_{\alpha}\,H\,\,,\rc\rc
R_{\widehat \alpha\,\,\widehat \beta}&=&-{1\over 2}\,
H^{-{8\over 3}}\,\partial_{\alpha}H\,\partial_{\beta}H\,+\,
{\delta_{\alpha\beta}\over 3}\,\Big[\,
H^{-{8\over 3}}\,\big(\,\partial_{\gamma}\,H\,\big)^2\,-\,
H^{-{5\over 3}}\,\partial^2_{\gamma}\,H\,\Big]\,\,.
\label{apanueve}
\eear

\vskip 1cm
\renewcommand{\theequation}{\rm{B}.\arabic{equation}}
\setcounter{equation}{0}
\medskip
\section{Solution of the profile equation}
\medskip
Suppose that $f(\,\vec x\,)$ is
a function depending on the $d$-dimensional vector $\vec x$ which satisfies the
equation:
\beq
\nabla^2_d\,f\,=\,{C\over |\,\vec x\,|^n}\,\,,
\label{apbuno}
\eeq
where $\nabla^2_d$ is the laplacian operator in $d$ dimensions and $C$ is a
constant. For $d\not =n$ we have the following solution of the above equation:
\beq
f(\,\vec x\,)\,=\,
\cases{{C\over (d-n)(2-n)}\,|\,\vec x\,|^{2-n}\,,
&\,\,\,\,\,$n\not= 2\,\,,$\cr\cr
{C\over 2(d-2)}\,\,\log  (\vec x^{\,2})\,\,,
&\,\,\,\,\,$n= 2\,\,.$}
\label{apbdos}
\eeq
Notice that the general solution of eq. (\ref{apbuno}) can be obtained by
adding
a $d$-dimensional harmonic function to the particular solution displayed
in eq. (\ref{apbdos}).

\vskip 1cm
\renewcommand{\theequation}{\rm{C}.\arabic{equation}}
\setcounter{equation}{0}
\medskip
\section{M-branes in the 20-SUSY pp-wave}
\medskip

In this appendix the general analysis given in (\ref{ttres}) will be
instrumental.  Hereafter, and in order not to clutter the notation, all
$\Gamma$ matrices will be flat by default and therefore hats will be
omitted everywhere. The easiest way to construct a pp-wave background that
leads to an enhancement of
$16$ to $20$ supersymmetries is to set one of the $\mu$'s, say $\mu_4$,
equal to zero in (\ref{siete}). Moreover, we will also take 
$\mu_1+\mu_2+\mu_3=0$, which ensures that the wave profile does not depend
on $y_9$.
This amounts to a coordinate split $x_i =
(y_1,y_1,y_3,y_4,y_5,y_6,z_7,z_8,y_9)$ with $y_i$($z_i$) tangent
(perpendicular) to the flux:
\beq
F_{wave}=dx^+\wedge\left(\mu_1 \,dy_1\wedge dy_2\wedge dy_9 +
\mu_2 \, dy_3\wedge dy_4\wedge dy_9 + \mu_3 \, dy_5\wedge dy_6\wedge dy_9
\rule{0mm}{5mm}\right)\,\,. \label{fcuawa}
\eeq
 Using now (\ref{siete}),  we obtain the
following undeformed profile:
$$
W_0 = -\frac{\mu_1^2}{4}( y_1^2+y_2^2) -\frac{\mu_2^2}{4}( y_3^2+y_4^2)
 -\frac{\mu_3^2}{4}( y_5^2+y_6^2)\,\,.
$$
As there are now three groups of $y_a$ coordinates,
namely $(y_1,y_2)$,  $(y_3,y_4)$ and $(y_5,y_6)$,  generic
M-brane embeddings will be labeled by $(+,-,(m_1,m_2,m_3),p,q)$
for a brane that extends along coordinates $(x^+,x^-)$,
$m_1$ out of $(y_1,y_2)$, $m_2$ inside $(y_3,y_4)$ and $m_3 $
along $(y_5,y_6)$, as well as  $p$ out of the $(z_7,z_8)$ and $q$
along $y_9$.  Clearly $m_1+m_2+m_3+p+q$ adds up to
$1$ for an $M2$ and to $4$ for an $M5$.

The 4-form flux  receives, as before, two contributions $F =
F_{wave} + F_{brane}$ coming from the wave and the brane respectively.
Typically it is the first piece that will signal the smearing
of the harmonic profile $H$ along directions perpendicular to the
embedding. This piece, as given in (\ref{fcuawa}), satisfies
the Bianchi identity $dF_{wave}= 0$  trivially. However from an analysis of
the  Maxwell's equations $d{}^* F_{wave}= 0$  the following conditions
are met for a coordinate $y_{\alpha}$ to be such that  $\partial_{\alpha} H
= 0$:

\bigskip
\centerline{
\begin{tabular}{|c|c|c|} \hline
$M2$ & $M5$ & smeared coordinates \\ \hline
$m_1+q \neq 0$ & $m_1+q-1 \neq 0 $ & $  y_1,y_2,y_9 $ \\
$m_2+q \neq 0$ & $m_2+q-1 \neq 0 $ & $  y_3,y_4,y_9 $ \\
$m_3+q \neq 0$ & $m_3+q-1 \neq 0 $ & $  y_5,y_6,y_9 $ \\
\hline
\end{tabular}
}

\bigskip

In the next paragraphs we shall investigate case by case all
possible embeddings. The full set of equations given
in  (\ref{ttres}) will be needed, as the possibility arises now
that $\theta'$ or $\theta''$ have zero modes, even if they
involve part of $\theta$. These cases will typically enforce
equality of two of the $\mu_\alpha$  (say $\mu_1 = \mu_2$).
In this sense it is worthwhile to remind the reader that
an equation like $\theta'\chi = 0$ with
$\theta' = \mu_1\Gamma_{129} + \mu_2\Gamma_{349} + \mu_3 \Gamma_{569}$
and $\mu_1+\mu_2+\mu_3=0$ is equivalent to a pair of projections, for example
$\Gamma_{1234}\chi = \chi~,~\Gamma_{1256}\chi = \chi$, and, therefore, it
enforces  a $1/4$ SUSY projection. However, if $\theta' =
\mu_1\Gamma_{129} +
\mu_2\Gamma_{349}$, only for $\mu_1= \mu_2$ there will be zero modes $\chi $
and the equation $\theta'\chi = 0$ will be equivalent to the
 single projection
$\Gamma_{1234}\chi =
\chi$.

In each case, the analysis will be divided in two parts. First,
the brane probe analysis will be carried out. Notice that, in general,
the Killing spinors for the supersymmetries of the brane probe
embeddings must be a subset of those corresponding to the  pure  pp-wave
configuration.  Hence in all cases we must have at least    the projection
$\Gamma_-\epsilon = 0$ for standard spinors, and $\Gamma_+\chi = \theta\chi
= 0$
for supernumeraries.

After that, the sugra analysis
can be performed. It involves in addition the following equations:
$$
\partial_\alpha \,\theta\,\epsilon~=~ \partial_\alpha \Omega_a\epsilon
~=~ \partial_{[\alpha}\Omega_{\beta]}\epsilon~ = ~0\,\,.
$$
For standard spinors, only the first one needs to be checked. Also it turns
out that the other two in general do not modify the brane-probe analysis
with excited scalars. Therefore, as a general rule, the only
effect of turning on the back-reaction is to kill
standard spinors in some cases.

In what follows we shall list the results of our analysis, both for the M2 and
M5 cases. For the sake of simplicity we will only  write down the
expressions of
$\theta'$ and $\theta''$ in the warped metric of the wave-brane background. The
corresponding values in the brane probe approach can be obtained by setting
to one the warp factors. We will also  indicate in each case the smearing
needed in the supergravity solution, the form of the harmonic function and the
profile of the brane-wave background.

\subsection{M2 branes}
\begin{itemize}
\item{\underline{$(+,-, ( 0,0,1),0,0)$}},
with the brane extending along $y_5$, and smeared along
$y_6,y_9$. In this case:
\beqa
\theta' &=& H^{-1/2}(\mu_1\Gamma_{129} +
\mu_2\Gamma_{349}) ~~~;~~~\theta''= \mu_3\Gamma_{569}\,\,,\nonumber\\
H &=& 1 +
\frac{Q}{\left( y_1^2+y_2^2+y_3^2+y_4^2+z_7^2+z_8^2\right)^2}\,\,,
\\ \nonumber
W &=& W_0
~+~\frac{1}{8}
\frac{Q \mu_3^2}{ y_1^2+y_2^2+y_3^2+y_4^2+z_7^2+z_8^2 }\,\,.
\eeqa
\begin{description}
\item{- Brane Probe}: Setting $Q=0$ in $H$ first, we see that, only if  $\mu_1=
\mu_2$ ,
$\theta'$ has a chance to have zero modes: $~\theta'\chi = 0 \Leftrightarrow
\Gamma_{1234}\chi = \chi$. In this case we find
4 standard spinors $\epsilon  =
e^{\frac{x^+}{4}\theta''}\chi$ with $ \Gamma_{+-5}\chi =
\chi\, ,
\theta'\chi = \Gamma_{-}
\chi = 0$ . For supernumeraries we obtain 0,
since in this case $\Omega_\alpha''  \sim
\Gamma_{-}\Gamma_{ \alpha}\theta''  $ has no zero modes.
Altogether this gives $4 (4+0)$ supersymmetries.
\item{- Sugra:}
The additional condition $\partial_\alpha\theta\epsilon = 0$ is met
iff $\theta'\epsilon=0$. So, we get the same set of projections for
standard spinors
$\eta = H^{-1/6}\epsilon= H^{-1/6}e^{\frac{x^+}{4}\theta''}\chi$. Since
there were already no supernumeraries, we get
$4 (4+0)$ supersymmetries.
\end{description}

\item {\underline{$(+,-, ( 0,0,0),1,0)$}}.
The brane extends along the $z_7$ coordinate. There is no smearing and
the profile exhibits no deformation:
\beqa
\theta' &=&
H^{-1/2}(\mu_1\Gamma_{129} +  \mu_2\Gamma_{349}
+\mu_3\Gamma_{569} ) ~~~;~~~\theta'' = 0\,\,,   \nonumber  \\
H &=& 1 + \frac{Q}{\left(
(\vec{\tilde y} )^2 + \tilde z_8^2
\right)^3}\,\,,
\\ \nonumber
W  &=& W_0\,\,.
\eeqa
\begin{description}
\item{- Brane Probe:} Setting $Q=0$ in $H$, we get  2 standard (constant)
spinors
$\epsilon =\chi$, with
$\theta'\chi =   \Gamma_{-}\chi = 0 $ and $\Gamma_{+-7}\chi = \chi$.
Also there are 2 supernumerary spinors $\epsilon =
e^{x^\alpha\Omega_\alpha}\chi$
with $\Omega_a\chi = \Omega_7\chi = 0 \Leftrightarrow \theta\chi =
0$, as well as $\Gamma_+\chi = 0$ and
$\Gamma_{+-7}\chi = \chi$.  We find altogether $4 (2+2)$ supersymmetries.
\item{- Sugra:} Here also, when $Q\neq 0$, the constraint
$\partial_\alpha\theta\epsilon = 0$ is fulfilled   if $\theta'\epsilon =
0$. Therefore, the same number of standard spinors as in the brane probe
approach occur and
$\eta=H^{-1/6}\epsilon = H^{-1/6}\chi$. Also, from
(\ref{stsiete}) we see that $\partial_\beta\Omega_\alpha = 0$,
and $\partial_\alpha\Omega_a\epsilon\sim \theta'\epsilon = 0$.
Thus,  two supernumerary spinors also survive:
$\eta=H^{-1/6}\epsilon=  H^{-1/6}e^{x^\alpha\Omega_\alpha}\chi$. Then,
we find  $4 (2+2)$ supersymmetries.
\end{description}
\item
{\underline{$(+,-, ( 0,0,0),0,1)$}}.
The brane extends along $y_9$ and   is smeared along
all $y_1,....,y_6$. Now:
\beqa
\theta' &=& 0
 ~~~;~~~\theta'' = \mu_1\Gamma_{129} +  \mu_2\Gamma_{349}
+\mu_3\Gamma_{569}\,\,,  \nonumber \\
H &=& 1 + Q \log (\,\vec z^{\,2\,}\,)\,\,, \\
W &=& W_0 - \frac{Q(\mu_1^2 + \mu_2^2 + \mu_3^2)}{4} ~\vec z^{\,2\,} ~
\left(\log(\,\vec z^{\,2\,}\,)-1\right)\,\,. \nonumber
\eeqa
\begin{description}
\item{- Brane Probe:} There are 8 standard spinors $\epsilon=
e^{\frac{x^+}{4}\theta''}\chi$ with $\Gamma_{+-9}\chi =\chi,~\Gamma_-\chi
= 0$. We have 2 supernumeraries without scalars $\epsilon=
e^{x^9\Omega_9}\chi$,
$\Gamma_{+-9}\chi =\chi,~\Gamma_+\chi =  \theta \chi= 0$.
With scalars, the condition $\Omega''_\alpha\chi = 0$ is impossible for
$\alpha=1,...,6$ and,
hence, there are no spinors. In summary we obtain $10 (8+2)$
supersymmetries without scalars and
$8 (8+0)$ with scalars.
\item{- Sugra:} If $Q\neq 0$, still $\partial_\alpha\theta=0$,
  so the brane-probe analysis is not modified for
standard spinors. For supernumeraries, $\partial_\beta\Omega_\alpha\neq 0$
(see (\ref{stsiete})), hence one must have  $\Omega_\alpha\epsilon=0$,
which is
again impossible. Therefore we find
$8(8+0)$ supersymmetries.
\end{description}

\end{itemize}

\subsection{M5 branes}

\begin{itemize}
\item \underline{$(+,-,(2,2,0),0,0)$}.
   The brane extends along $y_1,...,y_4$ and is smeared along $y_5,
y_6$ and $y_9$. In this case:
\beqa
\theta' &=& 0~~~~;~~~~\theta'' =\mu_1\Gamma_{129} +  \mu_2\Gamma_{349}\,\,,
+ H^{-1}\mu_3\Gamma_{569}\,\,, \nonumber \\
H &=& 1 + Q\log (\,\vec z^{\,2\,}\,) \,\,.\nonumber
\eeqa
\begin{description}
\item{- Brane Probe:} Setting $Q=0$   yields
 8 standard spinors    $\epsilon= e^{\frac{x^+}{4}\theta''}\chi$
with $\Gamma_{+-1234}\chi =\chi,~\Gamma_-\chi = 0$.
Also we get
 2 supernumerary spinors without scalars
$\epsilon= e^{x^a\Omega_a}\chi; $
$\Gamma_{+-1234}\chi =\chi,~\Gamma_+\chi = \theta \chi= 0 $.
With scalars, $\Omega''_\alpha\chi = 0$ is impossible for
$\alpha = y_5, y_6$, therefore  there are no supernumerary spinors in this
case.
In summary, we obtain $10 (8+2)$ and $8 (8+0)$ supersymmetries.

\item{- Sugra:}
Now, with $Q\neq 0$ the profile $W$ is difficult to
solve for. Moreover, we have the additional  integrability condition
$\partial_\alpha\theta\epsilon = 0$, which is impossible to fulfill because
$\partial_\alpha\theta\sim \Gamma_{569}$. Therefore, we do not have any
supersymmetry in this case.
\end{description}

\item \underline{$(+,-,(1,1,2),0,0)$}. The brane extends along $y_1,y_3,
y_5$ and $y_6$ and is smeared along $y_9$. One has:
\beqa
\theta' &=& H^{-1/2}(\mu_1\Gamma_{129} +
\mu_2\Gamma_{349})
~~~~;~~~~\theta'' =   \mu_3\Gamma_{569}\,\,, \nonumber \\
H &=& 1 +  \frac{Q}{y_2^2 + y_4^2 + \vec z^{\,2}} \,\,,\nonumber\\
W &=&  W_0 + \frac{Q(\mu_1^2+ \mu_2^2)}{8}
\log (\,y_1^2 + y_2^2 + \vec z^{\,2\,}\,)\,\,.
\nonumber
\eeqa
\begin{description}
\item{- Brane-Probe:} Setting $Q=0$, only for $\mu_1 = \mu_2$~,   four standard
spinors exist,
$\epsilon = e^{\frac{x^+}{4}\theta''}\chi $,  with $\Gamma_{+-1356}\chi =
\chi$  and
$\theta'\chi = \Gamma_-\chi = 0$. For supernumerary spinors without scalars we
must impose
$\Omega'_a \chi =  0 $. This can have a solution for $a= 5,6$ if $\mu_1=
\mu_2$,  but not for $a=1,3$.
Therefore we find no supernumerary spinors.
 \item{- Sugra:}
For $Q\neq 0$, the supergravity analysis coincides with the brane-probe
analysis because
$\partial_\alpha \theta \chi \sim \theta'\chi = 0$ is one of the
defining conditions of the standard spinors and, thus,  there are no
supernumeraries.  In all cases we find  $4 (4+0)$ supersymmetries.
\end{description}

\item \underline{$(+,-,(1,1,1),1,0)$}.  The brane extends along $y_1, y_3,
y_5$ and $z_7$ with no smearing and:
\beqa
\theta' &=& H^{-1/2}(\mu_1\Gamma_{129} +
\mu_2\Gamma_{349} +\mu_3\Gamma_{569})
~~~~;~~~~\theta'' =0\,\,, \nonumber \\
H &=& 1 +  \frac{Q}{(y_2^2 + y_4^2 + y_6^2 +   z_8^2+ y_9^2 )^{3/2}}\,\,,
\nonumber\\ W &=&  W_0 - \frac{1}{8}\frac{Q(\mu_1^2+ \mu_2^2+ \mu_3^2)}{(y_2^2
+ y_4^2 + y_6^2 +   z_8^2+ y_9^2 )^{1/2} } \,\,.
\nonumber
\eeqa
 \begin{description}
\item{- Brane Probe:} with $Q=0$ we find 2 standard spinors $\epsilon =
\chi~(\Gamma_{+-1357}\chi =
\chi, \Gamma_-\chi = \theta' \chi = 0)$. Also we find
 0 supernumerary spinors because $\Omega'_a\chi = 0$  for
$a=z_7$ implies $\theta'\chi = 0$, but then for $a=1,3,5$ there is no solution.
\item{- Sugra:} With $Q\neq 0$, the same projections
as in the supergravity analysis are obtained, namely
$\partial_\alpha\theta \epsilon = 0
\Leftrightarrow \theta'=0$.  Altogether
we have $2(2+0)$ supersymmetries in all cases.
\end{description}

\item \underline{$(+,-,(1,1,1),0,1)$}. Now, contrarily to the previous
case, smearing occurs along all the $y_\alpha$ directions. Therefore only
$z_7,z_8$ are  transverse and we find:
\beqa
\theta' &=& 0
~~~~;~~~~\theta'' = \mu_1\Gamma_{129} +
\mu_2\Gamma_{349} +\mu_3\Gamma_{569} \,\,, \nonumber \\
H &=& 1 + Q\log(z_7^2+  z_8^2 ) \,\,,
\nonumber\\ W &=&  W_0 + \frac{Q}{8}(\mu_1^2+\mu_2^2+\mu_3^2)\,\,\vec
z^{\,2}\,\log(\,\vec z\,^{\,2\,}-1)\,\,.
\nonumber
\eeqa
\begin{description}
\item{- Brane Probe:} Putting $Q=0$ in $H$,  8 standard spinors are
obtained: $\epsilon = e^{\frac{x^+}{4}\theta''}\chi
~(\Gamma_{+-1359}\chi = \chi, \, \Gamma_-\chi = 0)$. Without scalars,
 we find  2 supernumerary spinors  $\epsilon =
e^{x^a\Omega_a}\chi~(\Gamma_{+-1359}\chi = \chi, \Gamma_+\chi =
\theta\chi = 0)$. When scalars are excited, $\Omega''_\alpha \chi = 0$ is
impossible for $\alpha = 2,4,6$. Altogether for this configuration we get $10
(8+2)$ and $8 (8+0)$ supersymmetries.
\item {- Sugra:} If $Q\neq 0$ no change occurs since $\theta$ is
independent of $H$, and there are no supernumerary spinors. So,
in this case one also gets $8 (8+0)$ supersymmetries.
\end{description}

\item \underline{$(+,-,(2,1,0),1,0)$}. The harmonic function can only
depend on $y_3$ and $z_8$, all other directions being either world-volume
or smeared. Thus:
\beqa
\theta' &=& H^{-1/2}\mu_2 \Gamma_{349}
~~~~;~~~~\theta'' = \mu_1\Gamma_{129}
  + H^{-1}\mu_3\Gamma_{569}  \,\,,\nonumber \\
H &=& 1 + Q\log(y_3^2+  z_8^2 )\,\,.
\nonumber
\eeqa
The profile is difficult to solve for.  In any case the embedding is not
supersymmetric, since  $\theta'$ in this case has no zero modes.

\item \underline{$(+,-,(2,1,0),0,1)$}. In this case:
\beqa
\theta' &=& H^{1/2}   \mu_1\Gamma_{129} +
  + H^{-1/2}\mu_3\Gamma_{569}
~~~~;~~~~\theta'' = \mu_2 \Gamma_{349}\,\,.
\nonumber
\eeqa
\begin{description}
\item{- Brane Probe:} When $H=1$ and   $\mu_1 = \mu_2$ there are 4 standard
spinors
$\epsilon = e^{\frac{x^+}{4}\theta''}\chi~(\Gamma_{+-1239}\chi =
\chi,
\Gamma_-\chi = \theta'\chi = 0)$. Concerning supernumerary spinors,
they must satisfy $\Omega''_\alpha \chi = 0$. For $\alpha=7,8$
this is tantamount to $\theta''\chi = 0$, which is impossible, since
$\theta''$ has no zero modes. Hence no supernumerary spinors survive
and we have $4 (4+0)$ supersymmetries.
\item{-  Sugra:} From our general rule, the brane should be smeared in the
$y_4$ coordinate. However $F_{wave}\wedge F_{brane}$ is not zero unless the
brane is completely smeared and $H$ is constant. Thus, in this case the only
supergravity solution we find is the original pp-wave.

\end{description}
\item \underline{$(+,-,(2,0,0),2,0)$}. There is smearing along all
$y_\alpha$ coordinates. And since $z_a$ are also internal, there
is no external volume to the brane, and it dissolves completely,
reverting to the original $pp$-wave.

\item \underline{$(+,-,(2,0,0),1,1)$}. For this configuration:
\beqa
\theta' &=& H^{1/2} \mu_1\Gamma_{129} + H^{-1/2}(
\mu_2\Gamma_{349} +\mu_3\Gamma_{569})
~~~~;~~~~\theta'' = 0 \nonumber \,\,.
\nonumber
\eeqa
\begin{description}
\item{- Brane Probe:} When $H=1$ we find  2 standard spinors $\epsilon =
\chi~(\Gamma_{+-1279}\chi =
\chi$, and $\Gamma_-\chi = \theta'\chi = 0)$ and
 0 supernumerary spinors, since $\Omega'_a\chi=0$ is not possible with
$a=1,2$. Thus this system is $2 (2+0)$ supersymmetric.
\item{- Sugra:} In this case $d{}^*F=0$ without any smearing. However,
$F_{wave}\wedge F_{brane}$ is zero only when $H$ is constant, which corresponds
to the pure pp-wave.

\end{description}
\item
\underline{$(+,-,(1,1,0),2,0)$}. The brane extends along $y_1,y_3,z_7,
z_8$ and is smeared along $y_5, y_6, y_9$.  Now:
\beqa
\theta' &=& H^{-1/2} (\mu_1\Gamma_{129} +
\mu_2\Gamma_{349})
~~~~;~~~~\theta'' =  H^{-1} \mu_3\Gamma_{569} \,\,,
\nonumber \\
H &=&  1 + Q \log(\,y_2^2 + y_4^2\,)\,\,.
\nonumber
\eeqa
\begin{description}
\item{- Brane Probe:} with $Q=0$, only for $\mu_1 = \mu_2$ there are 4
standard spinors,
$\epsilon=e^{\frac{x^+}{4}\theta''}\chi~(\Gamma_{+-1378}\chi =
\chi, \Gamma_-\chi = \theta'\chi = 0)$, and
  0 supernumeraries, because $\theta''$ has no zero modes:
\ie\ this configuration is $4 (4+0)$ supersymmetric.
\item{- Sugra:}  with $Q\neq 0$ all spinors are lost since
$\partial_\alpha
\theta\epsilon= 0$ has no solution.
\end{description}
\item
\underline{$(+,-,(1,1,0),1,1)$}. The brane extends along $y_1,y_2,z_7, y_9$
and is smeared along $y_2,y_4$. Thus:
\beqa
\theta' &=& H^{-1} \mu
_3\Gamma_{569}
~~~~;~~~~\theta'' = \mu_1\Gamma_{129}  + \mu_2 \Gamma_{349}\,\,, \nonumber \\
H &=&  1 + \frac{Q}{(  y_5^2+  y_6^2 + z_8^2)^{1/2}}\,\,.
\nonumber
\eeqa
No spinors in any case, since $\theta'$ has no zero modes.
Profile $W$ seems  difficult to solve for.

\item \underline{$(+,-,(1,0,0),2,1)$}. The brane covers $y_1,z_7,z_8,y_9$
and is smeared along $y_2$. Therefore:
\beqa
\theta' &=& H^{-1/2} (\mu_2\Gamma_{349} +
\mu_3\Gamma_{569})
~~~~;~~~~\theta'' =    \mu_1\Gamma_{129} \,\,,
\nonumber \\
H &=&  1 + \frac{Q}{y_3^2+y_4^2+y_5^2+y_6^2}\,\,,
\nonumber\\
W &=& W_0 +
\frac{Q\mu_1^2}{8}\log(\,y_3^2+y_4^2+y_5^2+y_6^2\,)\,\,.
\nonumber
\eeqa

\begin{description}
\item{- Brane Probe:} When $Q=0$, only for $\mu_1=\mu_2$ there are $4$
spinors, $\epsilon= e^{\frac{x^+}{4}\theta''}\chi$, with
$\Gamma_{+-1789}\chi = \chi$ and $\Gamma_- \chi = \theta' \chi = 0$.
No supernumeraries appear because $\theta''$ has no zero modes.
\item{- Sugra:} For $Q\neq 0$,  the condition $\partial_\alpha\theta\chi =
0$ is fulfilled with $\theta'\chi = 0$. So, the analysis goes through, and
we have 4 standard spinors $\eta=H^{-1/12}\epsilon = H^{-1/12}
e^{\frac{x^+}{4}\theta''}\chi$,  with the same $\chi$ as before.
In all cases we have $4(4+0)$ supersymmtries.
\end{description}

\end{itemize}

The previous analysis is summarized in the following table, where we have
included only those cases which preserve some supersymmetry. The asterisk
distinguishes those configurations for which $\mu_1=\mu_2$ has to be
enforced in
order to have some supersymmetry.

\bigskip
\centerline{
\begin{tabular}{|c|c|c|c|}\hline
~ &  brane probe & brane probe & sugra \\
$M2$ &  without scalars & with scalars
& analysis \\ \hline
$(+,-, ( 0,0,1),0,0)^*$ &  4(4+0) & 4(4+0) & 4(4+0)
\rule{0mm}{5mm}\\
$(+,-, ( 0,0,0),1,0)$ &  4(2+2) & 4(2+2) & 4(2+2)
\rule{0mm}{5mm}\\
$(+,-, ( 0,0,0),0,1)$ &  10(8+2) & 8(8+0) & 8(8+0)
\rule{0mm}{5mm}\\
\hline
$M5$ & & &
\rule{0mm}{5mm}\\ \hline
 $(+,-, ( 2,2,0),0,0)$ &  10(8+2) & 8(8+0) & 0(0+0)
\rule{0mm}{5mm}\\
 $(+,-, ( 1,1,2),0,0)^*$ &  4(4+0) & 4(4+0) & 4(4+0)
\rule{0mm}{5mm}\\
 $(+,-, ( 1,1,1),1,0)$ &  2(2+0) & 2(2+0) & 2(2+0)
\rule{0mm}{5mm}\\
 $(+,-, ( 1,1,1),0,1)$ &  10(8+2) & 8(8+0) & 8(8+0)
\rule{0mm}{5mm}\\
 $(+,-, ( 2,1,0),0,1)^*$ &  4(4+0) & 4(4+0) & -
\rule{0mm}{5mm}\\
 $(+,-, ( 2,0,0),1,1)$ &  2(2+0) & 2(2+0) & -
\rule{0mm}{5mm}\\
$(+,-, ( 1,1,0),2,0)^*$ &  4(4+0) & 4(4+0) & 0(0+0)
\rule{0mm}{5mm}\\
$(+,-, ( 1,0,0),2,1)^*$ &  4(4+0) & 4(4+0) & 4(4+0)
\rule{0mm}{5mm}\\
\hline
\end{tabular}
}

\end{document}